\documentclass[11pt,letterpaper]{article}
\pdfoutput = 1
\usepackage     {jheppub}
\usepackage     {journals}
\usepackage     {verbatim}
\usepackage     [mathscr]{eucal}
\usepackage     {bm}
\bibpunct	{[}{]}{,}{n}{}{;}
\usepackage[utf8]{inputenc}

\def\tocsqueeze	{\vspace*{-3.7pt}}

\title          { On the Evolution of Jet Energy and Opening Angle  in Strongly Coupled Plasma }
\author[a]      {Paul~M.~Chesler,}
\author[b]	{Krishna~Rajagopal,}
\affiliation[a]	{Department of Physics, Harvard University, Cambridge MA 02138, USA}
\affiliation[b] {Center for Theoretical Physics, MIT,
                Cambridge MA 02139, USA}
\emailAdd	{pchesler@physics.harvard.edu}
\emailAdd       {krishna@mit.edu}

\preprint{MIT-CTP-4735}

\abstract
    {%
    We calculate how the energy and the opening angle of jets in ${\cal N}=4$ SYM theory
    evolve as they propagate through the strongly coupled plasma of that theory.  We define
    the rate of energy loss $dE_{\rm jet}/dx$ and the jet opening angle in a straightforward fashion directly in the gauge theory
    before calculating both holographically, in the dual gravitational description.    In this way, we rederive
    the previously known result for $dE_{\rm jet}/dx$ without the need to introduce a finite slab of plasma. 
    We obtain a striking relationship between the initial opening angle of the jet, which is to say the opening
    angle that it would have had if it had found itself in vacuum instead of in plasma, and the thermalization
    distance of the jet. Via this relationship, we show that ${\cal N}=4$ SYM jets with any initial energy that have the same initial opening
    angle and the same trajectory through the plasma experience the same fractional energy loss.  We also provide an expansion
    that describes how the opening angle of the ${\cal N}=4$ SYM jets increases slowly as they lose energy, over the fraction of their lifetime
    when their fractional energy loss is not yet large. We close by looking ahead toward potential qualitative lessons from our results for QCD jets
    produced in heavy collisions and propagating through quark-gluon plasma.
        }

\keywords	{gauge-gravity correspondence, 
		 quark-gluon plasma, jets, heavy ion collisions}
\begin{document}

\advance\textheight 55pt
\maketitle
\thispagestyle{empty}
\advance\textheight -55pt

\addtocontents	{toc}{\tocsqueeze}
\addtocontents	{toc}{\tocsqueeze}

\section{Introduction and summary of results}
\label{sec:Introduction}

Consider a light quark jet moving through 
an infinite volume of static strongly coupled plasma with temperature $T$.
Broadly speaking, the jet consists of an energetic spray of excitations 
localized within a narrow cone of opening angle $\theta_{\rm jet}$ about the axis of propagation of the jet.
As the jet propagates though the plasma, energy and momentum leave the jet and excite
hydrodynamic modes in the plasma.  The excited hydrodynamic modes
then transport the lost energy and momentum away from the jet.  
Interesting questions to consider include: what is the rate at which the jet loses energy to the plasma?  
How does $\theta_{\rm jet}$ change as the jet propagates? 
How far can the jet travel before it thermalizes?  
How do the rate of energy loss and the thermalization distance $x_{\rm therm}$
depend on initial conditions, including in particular the initial energy of the jet and
the initial value of $\theta_{\rm jet}$? 

Holographic duality \cite{Maldacena:1997re,Witten:1998qj,Gubser:1998bc} relates the dynamics of  
the strongly coupled plasma
in
certain large $N_{\rm c}$ 
gauge theories to the dynamics of classical black holes 
in one higher dimension.  The simplest theory 
with a dual gravitational description is 3+1-dimensional $\mathcal N = 4$ supersymmetric 
Yang-Mills theory (SYM), whose dual description 
is that of gravity in an asymptotically AdS$_5$ spacetime.  Thanks to this dual classical gravity description,
the properties of ${\cal N}=4$ SYM plasma are under much better theoretical control than is the case for QCD plasma.
This has motivated much interest in using ${\cal N}=4$ SYM plasma as a toy model for the 
formation, properties, and dynamics of real quark-gluon 
plasma and the dynamics of probes therein. (For reviews, see for example 
Refs.~\cite{Gubser:2009md,CasalderreySolana:2011us,Chernicoff:2011xv,DeWolfe:2013cua,Chesler:2015lsa}.)

Models for jet quenching  in the  strongly coupled plasma of  ${\cal N}=4$ SYM theory have been studied extensively. 
(See, for example, 
Refs.~\cite{Herzog:2006gh,%
        CasalderreySolana:2006rq,%
        Liu:2006ug,%
        Gubser:2006bz,%
        Chernicoff:2006hi,%
        Chesler:2007an,Gubser:2007ga,Chesler:2007sv,
        Gubser:2008as,%
               Chernicoff:2008sa,%
        Chesler:2008uy,%
        Arnold:2010ir,Arnold:2011qi,Arnold:2011vv,
        Chesler:2011nc,%
        Ficnar:2012np,
        Arnold:2012uc,Arnold:2012qg,
        Ficnar:2013wba,%
        Ficnar:2013qxa,%
        Chesler:2013cqa,%
        Chesler:2014jva,%
        Casalderrey-Solana:2014bpa,%
        Rougemont:2015wca,
        Casalderrey-Solana:2015vaa}.)
On first hearing this sounds surprising since, unlike in QCD, in ${\cal N}=4$ SYM theory hard processes with large momentum transfer do not
produce jets~\cite{Hatta:2008tx,Hofman:2008ar}.  Nevertheless, one can construct states in ${\cal N}=4$ SYM theory which have highly energetic 
and localized excitations --- which we shall call ``jets" --- which propagate arbitrarily far through plasma before thermalizing.  
Indeed, over the past decade, different authors~\cite{Herzog:2006gh,%
        CasalderreySolana:2006rq,%
        Gubser:2006bz,%
        Gubser:2008as,%
        Chesler:2008wd,
        Chesler:2008uy,%
        Arnold:2010ir,Arnold:2011qi,Arnold:2011vv,
        Chesler:2011nc,%
        Arnold:2012uc,Arnold:2012qg,
        Ficnar:2013wba,%
        Ficnar:2013qxa,%
                Chesler:2014jva%
        }
have introduced different ways of constructing states  in ${\cal N}=4$ SYM theory with ``jets." 
However, thus far most authors have defined physical quantities, such as the ``jet" energy loss rate, via gravitational quantities
instead of directly in terms of field theoretic variables.  Together with the differing models
for ``jets," this has led to different authors predicting qualitatively different energy loss rates \cite{Ficnar:2012np}.
Our strategy is to work the other way around.  We use the gravitational description 
to compute the expectation value of the ${\cal N}=4$ SYM stress tensor, $\langle T^{\mu \nu} \rangle$,
and extract physical quantities, including the rate of energy loss and the evolution of the 
opening angle, directly from $\langle T^{\mu \nu} \rangle$.

In this paper, we shall study the ``jets" introduced in Refs.~\cite{Chesler:2008wd,Chesler:2008uy}
and analyzed further in Ref.~\cite{Chesler:2014jva} which consist of massless quark jets
represented in the dual gravitational description as arcs of rapidly moving, falling, open strings.
We consider the propagation of these ``jets'' in an infinite volume of strongly coupled plasma,
dispensing with the finite slab of plasma introduced as a device in Ref.~\cite{Chesler:2014jva}.
There are (at least) two ways to use results about how ``jets" lose energy 
to gain insights into jet quenching in QCD.  
One option is to take the form for the ``jet" energy loss rate 
and apply it parton-by-parton to every parton in a QCD
jet, letting QCD describe the production of a hard parton and its subsequent fragmentation into
a shower, and using the ${\cal N}=4$ SYM ``jet" energy loss rate to describe how each parton in the shower
loses energy.  This yields the hybrid model for jet quenching introduced in Refs.~\cite{Casalderrey-Solana:2014bpa,Casalderrey-Solana:2015vaa}.
These authors have shown that this approach provides a good representation of much
existing jet data, and have used it to make predictions for many experimental measurements
soon to come.  The second possible path toward insights about jet quenching in QCD is more
radical: one can think of the ``jets" that we analyze literally as proxies for QCD jets, and look for
insights from  our calculation
into how jets change in shape, as well as lose energy, as they propagate through strongly
coupled plasma.  We shall take some steps in this second direction in Section~\ref{sec:concludingremarks}.
For this reason, and for simplicity, we shall henceforth refer to our ``jets'' just as jets.

We focus on jets that have vanishingly small size at the moment when they are produced.
In the dual gravitational description, this requires considering strings that are created at a point asymptotically close
to the AdS boundary.  
The jets we study initially expand linearly in time as they propagate, meaning that initially
they expand with a constant opening angle $\theta_{\rm jet}$. 
If the jets were in fact in vacuum,  $\theta_{\rm jet}$ would remain constant at
its initial value $\theta_{\rm jet}^{\rm init}$.
For the jets we consider, though,
as they propagate their opening angle $\theta_{\rm jet}$ increases as
they lose energy and excite hydrodynamic modes of the plasma. 
If the volume of plasma in which they are propagating is large enough, eventually after traveling a distance $x_{\rm therm}$  they
have lost all of their initial energy and they thermalize.  We shall see that $x_{\rm therm}T$ is determined by $\theta_{\rm jet}^{\rm init}$.
In any consideration of high energy jets produced in heavy ion collisions, the plasma will end
before the jet thermalizes, meaning that jets emerge from it rather than thermalizing in it.
We shall see, though, that the thermalization distance
$x_{\rm therm}$ plays a central role throughout our calculations, 
including in the expressions 
we obtain that describe how the energy and opening angle of jets that travel only a fraction of $x_{\rm therm}$ evolve.

In order to crisply separate the jet from the plasma and thereby systematically define the 
properties of the jet, including its rate of energy loss  and its opening angle, it is useful to 
focus on states where there is a large separation of scales.  Indeed, as $1/T$ is the characteristic 
microscopic scale in strongly coupled plasma, it makes no sense to attempt to 
define instantaneous properties of a jet with any spatial or temporal fidelity smaller than $1/T$! 
With this in mind, we study jets whose $x_{\rm therm} \ggg 1/T$ and focus on physics in the region 
\begin{align}
\label{SteadyStateRange}
&&x \gg \frac{1}{T},&
&x_{\rm therm} - x \gg \frac{1}{T},&&
\end{align}
which we shall refer to as the ``steady-state region" (SSR).  Jets whose $x_{\rm therm} \ggg 1/T$
have high initial energy $E_{\rm init}\ggg T\sqrt{\lambda}$ where $\lambda$ is the (large) 't~Hooft coupling of
the strongly coupled gauge theory \cite{Chesler:2008uy,Gubser:2008as}.
The SSR 
becomes arbitrarily close to 100\% of the jet's trajectory as $x_{\rm therm} T \to \infty$,
since the SSR includes any $x$ such that $x/x_{\rm therm}$ tends to
a constant between 0 and 1 in this limit. 
In the SSR, as the jet moves over distances $\ell$ with $1/T \ll \ell \ll x_{\rm therm}$, the evolution of the
jet is steady-state, 
with deviations from steady-state behavior suppressed by powers of $ \ell/x_{\rm therm}$. 
This allows for simple definitions of both the instantaneous rate of energy loss 
$d E_{\rm jet}(x)/dx$
and the instantaneous opening angle $\theta_{\rm jet}(x)$ in terms of $\langle T^{\mu \nu} \rangle$, 
definitions which we give below in Sec.~\ref{sec:elossdef}.  
As we shall see below, as $x \to x_{\rm therm}$ and the jet exits the SSR, the energy loss rate and opening angle grow rapidly, 
the separation between the jet and plasma becomes less and less distinct, 
and the jet's instantaneous energy loss rate and opening angle become less and less crisply defined.

We now summarize our principal results.  

We find that jets with thermalization distance $x_{\rm therm}\ggg 1/T$ 
have initial opening angles 
\begin{equation}
\label{eq:openingangle}
\theta_{\rm jet}^{\rm init} = \kappa   \left ( \frac{E_{\rm init}}{E_0} \right )^{-2/3},
\end{equation} 
where $E_{\rm init}$ is the initial jet energy, $E_0$ is an energy scale
that depends on how the jet is prepared,
and $\kappa$ is an ${\cal O}(1)$ pure number. We shall discuss each of these three quantities in turn.

The initial jet energy $E_{\rm init}$ in (\ref{eq:openingangle}) is the energy when the jet 
enters the SSR. It does not make sense to define the jet energy
until some time $\gtrsim 1/T$ after its creation, as 
at its creation gluon fields around its creation event are created and we need to wait for the jet to
separate from gluon fields that are not part of the jet before defining the energy of the jet.  This is equally true in vacuum
or in plasma, and it is somewhat analogous to what is described by soft functions in jet production calculations in QCD.
In the strongly coupled plasma, the gluonic energy that is not co-moving with the jet thermalizes within a time of order $1/T$,
after which the initial jet energy $E_{\rm init}$ can be crisply defined.
The constant $E_0$ is a temperature-independent energy scale proportional to $\sqrt{\lambda}$ with
a temperature-dependent minimum value proportional to 
$T\sqrt{\lambda}$, meaning that $E_0$ must be large compared to $T$.  $E_0$ is small compared to $E_{\rm init}$ so $\theta_{\rm jet}^{\rm init} \ll 1$.
The value of $E_0$ depends on details of how the state is prepared.
The constant $\kappa$ appearing in (\ref{eq:openingangle}) depends on the precise definition of the jet opening angle,
but once this is defined, $\kappa$ is the same for every jet.  We shall define $\theta_{\rm jet}$ precisely
in Sec.~\ref{sec:elossdef}, but it is in essence the half-width-at-half-maximum of the jet, meaning that the flux
of energy an angle $\theta_{\rm jet}$ away from the jet axis is half as large as the flux along the jet axis.  
With our definition of  $\theta_{\rm jet}$,
we find $\kappa \approx 0.204157$. (If, for example, we had instead defined $\theta_{\rm jet}$
as the angle at which the energy flux is 10\% of its maximum value on the jet axis,
$\kappa$ would be larger by a factor of 2.23507 and the angular size each of our jets, defined
in this alternate fashion, would be
larger by this factor than is the case with the definition we have chosen.)


Likewise, the thermalization distance reads ~\cite{Chesler:2008uy,Gubser:2008as}
\begin{equation}
\label{eq:thermalizationdistance0}
x_{\rm therm} = \frac{1}{T} \left ( \frac{E_{\rm init}}{E_0} \right )^{1/3} \,,
\end{equation}
for the same $E_0$.
Therefore, among the jets with some given initial energy $E_{\rm init}$, the jets that have the maximum
possible the thermalization distance (\textit{i.e.} have the minimum possible value of $E_0$)
have a minimal but nonzero opening angle.  
To make contact with notation introduced in previous literature, we can introduce the dimensionless constant $\mathcal C$ via 
$E_0= T\sqrt{\lambda}/{\mathcal C}^3$, so that Eq.~(\ref{eq:thermalizationdistance0}) takes the form
$x_{\rm therm} = \frac{\mathcal C}{T} \left ( \frac{E_{\rm jet}}{T \sqrt{\lambda}} \right )^{1/3}$.  The maximum 
value of $\mathcal C$, corresponding to the minimum value of $E_0$, 
has been estimated to be around 0.5 to 1 \cite{Chesler:2008uy,Ficnar:2013wba}.
We see from (\ref{eq:openingangle}) that
the fact that $E_0$ has a minimum value means that, for a given $E_{\rm init}$, the 
initial opening angle $\theta_{\rm jet}^{\rm init}$ has a nonzero minimum value. 
Equivalently, it means that the range of possible values of $E_{\rm init}$ for jets with a given $\theta_{\rm jet}^{\rm init}$
has a nonzero minimum value.

Using (\ref{eq:openingangle}), $E_0$ can be eliminated from (\ref{eq:thermalizationdistance0}) 
to yield
\begin{equation}
\label{eq:xtherm2}
x_{\rm therm} = \frac{1}{ T} \sqrt{\frac{\kappa}{\theta^{\rm init}_{\rm jet}}}.
\end{equation}
We therefore see that the thermalization distance is entirely determined by the initial opening angle
of the jet and the temperature of the plasma; it is independent of the details of how the jet is prepared
that are encoded in $E_0$.\footnote{This independence of the thermalization
distance on extraneous details is obtained for jets that have zero size 
when they are produced and that initially behave as if they were produced in vacuum. We are considering
only such jets. They
expand
with an opening angle that is initially $\theta_{\rm jet}^{\rm init}\neq 0$, as if in vacuum, 
and subsequently increases due
to the presence of the plasma, as we shall see below.
If, as in Refs.~\cite{Ficnar:2013wba,Ficnar:2013qxa}, one instead considers jets
produced with a nonzero initial size, the relation (\ref{eq:xtherm2}) need not be valid.  For example, it is possible 
(although doing so
takes a long time $\ggg 1/T$ in the past, arranging the shape of the trailing string)
to prepare `initial' states with nonzero size that have $\theta_{\rm init}=0$ but have
a finite $x_{\rm therm}$~\cite{Ficnar:2013wba,Ficnar:2013qxa}.
}
This is a remarkable result, as it says that all jets with a given initial opening angle
$\theta_{\rm jet}^{\rm init}$, with any value of the initial energy above the minimum possible,
have the same thermalization distance.

\begin{figure}
\begin{centering}
\includegraphics[width=14cm]{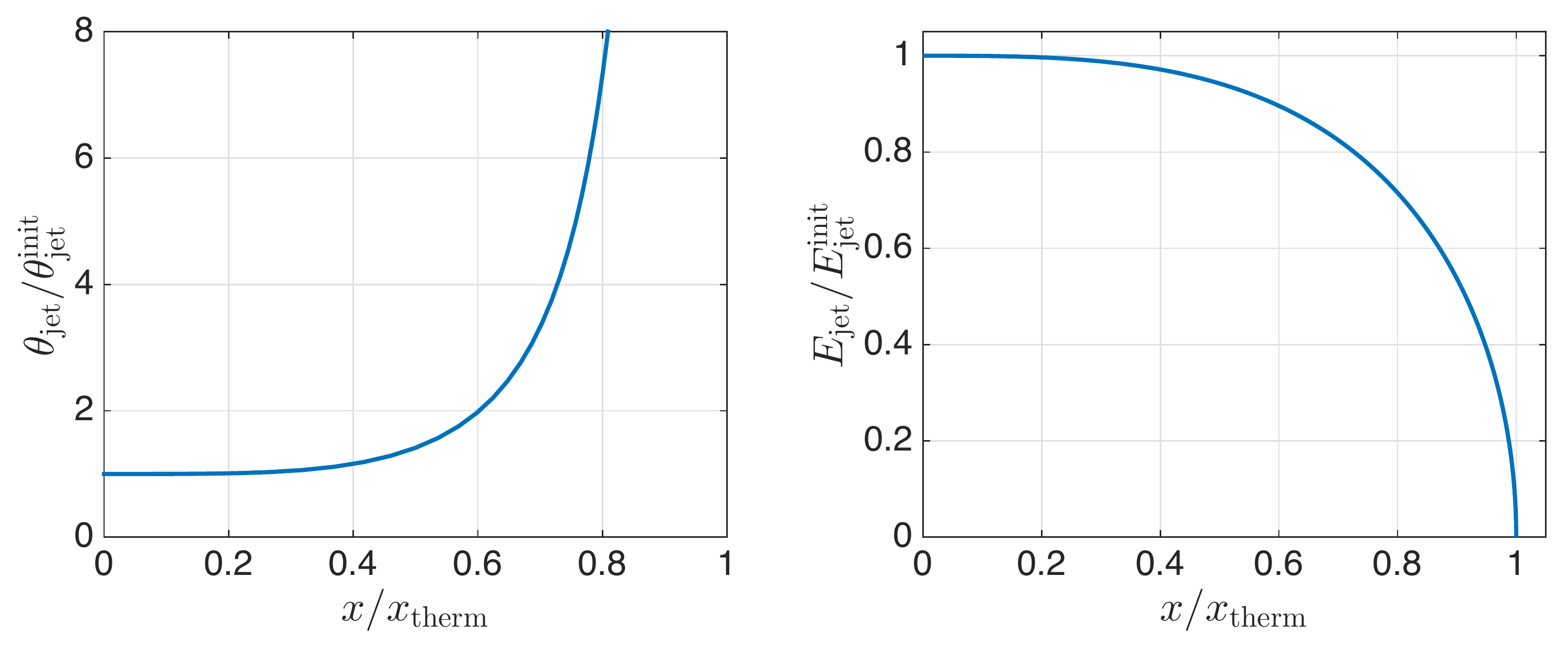} 
\par\end{centering} 
\protect\caption{
\label{fig:IntroResults} 
Left: the opening angle of the jet, $\theta_{\rm jet}$, normalized by
its initial value 
as a function of $x/x_{\rm therm}$, as given in Eq.~(\ref{eq:mainopeningangleresult}).  
The opening angle increases slowly 
until $x/x_{\rm therm} \sim 0.5$.
As $x$ increases further, $\theta_{\rm jet}$ increases like 
$\theta_{\rm jet} \sim \frac{\theta_{\rm jet}^{\rm init}}{\left [1 - x/x_{\rm therm} \right ]^2}.$
As $x \to x_{\rm therm}$ the jet exits the SSR, the instantaneous opening angle becomes 
ill-defined, and the jet thermalizes.  
When $x_{\rm therm} - x \sim 1/T$, the
jet opening angle has grown to 
$\theta_{\rm jet} = {\cal O}(1)$, meaning that the jet thermalizes when its opening angle becomes of 
order 1.  Our results are only valid in the SSR: our calculations break down where $x_{\rm therm} - x \lesssim 1/T$.
This means that $\theta_{\rm jet}$ is only described by the rapidly rising curve in the figure until
it reaches ${\cal O}(1)$.  If its initial value $\theta_{\rm jet}^{\rm init}$ is very small, this happens
when $x/x_{\rm therm}$ is close to 1.
Right: The jet energy $E_{\rm jet}$, normalized by its initial value, as a function of $x/x_{\rm therm}$, as given 
by integrating Eq.~(\ref{eq:energylossresult0}).
The jet loses energy slowly until $x/x_{\rm therm} \sim 0.5$.  The jet energy has 
decreased by 30\% at $x/x_{\rm therm} = 0.8$ and by 83\% at $x/x_{\rm therm} = 0.99$. 
}
\end{figure}

We find that in the SSR the jet's instantaneous opening angle
increases steadily as the jet propagates through the strongly coupled plasma
and is well-approximated by 
\begin{equation}
\label{eq:mainopeningangleresult}
\frac{\theta_{\rm jet}(x)}{\theta_{\rm jet}^{\rm init}} \approx \sqrt{1 + \left[ F^{-1}\left( \frac{\textstyle x}{\textstyle x_{\rm therm}}\right)\right]^4},
\end{equation}
with $F^{-1}$ the inverse of 
\begin{equation}
\label{eq:Fdef2}
F(\hat u) \equiv  1 - { \frac{4 \sqrt{\pi}}{ \Gamma\left( {\textstyle \frac 14} \right)^2} } \,  \frac{ {_2F_1}\left({\textstyle \frac 14, \frac 12, \frac 54,-\frac{1}{\hat u^4}}\right)}{\hat u}    
\,,
\end{equation}
with ${_2F_1}$ the Gauss Hypergeometric function.  
The approximation (\ref{eq:mainopeningangleresult}) describes the full result obtained
in Sec.~\ref{sec:openingangle}
to within 1.5\%.
In the left panel of Fig.~\ref{fig:IntroResults}, we plot the approximation (\ref{eq:mainopeningangleresult})  
for $\theta_{\rm jet}/\theta_{\rm jet}^{\rm init}$. 
We see from Fig.~\ref{fig:IntroResults} that 
the opening angle increases only very slowly
until $x/x_{\rm therm} \sim 0.5$.  Indeed, at small $x/x_{\rm therm}$ we can expand Eq.~(\ref{eq:mainopeningangleresult}), obtaining
\begin{equation}
\label{eq:jetangleresultSmallx}
\frac{\theta_{\rm jet}}{\theta_{\rm jet}^{\rm init}}  \approx  1 + 2 \left( \frac{x}{a\,x_{\rm therm}} \right)^4 + \frac{6}{5}\left( \frac{x}{a\,x_{\rm therm}} \right)^8
+ \frac{44}{75}\left( \frac{x}{a\,x_{\rm therm}} \right)^{12}
+ \ldots\ ,
\end{equation}
with the constant
\begin{equation}
\label{eq:adef}
a\equiv  \frac{4\sqrt{2\pi}}{\Gamma\left(\frac{1}{4}\right)^2} \approx 0.763\,.
\end{equation}
In any consideration of jets in heavy ion collisions which travel through a length of plasma that
is much less than $x_{\rm therm}$, the expansion (\ref{eq:jetangleresultSmallx}) is
the form of our result for the evolution of the jet opening angle that is of interest.
We can, however, use the approximation (\ref{eq:mainopeningangleresult}) at larger values of $x$.
It is easy to see from (\ref{eq:mainopeningangleresult}) that as $x \to x_{\rm therm}$ the opening angle grows like
\begin{equation}
\label{eq:thetadivergence}
\theta_{\rm jet} \sim \frac{\theta_{\rm jet}^{\rm init}}{\left[ 1 - \frac{{\textstyle x}}{\textstyle{x_{\rm therm}}}\right]^2}.
\end{equation}
From this expression and (\ref{eq:xtherm2}), we see that as $x$ increases to 
$x_{\rm therm} - x \sim 1/T$ --- where the jet exits the SSR --- the opening angle of the jet increases rapidly to 
$\theta_{\rm jet}={\cal O}(1)$.  At this point neither our result (\ref{eq:mainopeningangleresult}), plotted in the left
panel of Fig.~\ref{fig:IntroResults}, nor the expression (\ref{eq:thetadivergence}) are valid any longer.
In fact, once $x_{\rm therm}-x \sim 1/T$ the jet is itself no longer sharply defined.


We also calculate the instantaneous rate of energy loss for the jet while it is in the SSR and find
\begin{equation}
\label{eq:energylossresult0}
\frac{1}{E_{\rm init}} \frac{dE_{\rm jet}}{dx} = - \frac{4 x^2}{\pi x_{\rm therm}^2 \sqrt{x_{\rm therm}^2 - x^2}}.
\end{equation}
This expression is identical to that obtained in our previous work~\cite{Chesler:2014jva}.  
From Eqs.~(\ref{eq:xtherm2}) and (\ref{eq:energylossresult0}) we therefore see that the energy loss rate 
is, up to normalization, entirely fixed by the opening angle of the jet and the plasma temperature.
In particular, the {\it fractional} energy loss $\Delta E_{\rm jet}/E_{\rm init}$ suffered by a jet propagating for a distance
$x$, obtained by integrating (\ref{eq:energylossresult0}), is entirely determined by $x$, $T$ and $\theta_{\rm jet}^{\rm init}$
in the single combination $x/x_{\rm therm}$ and doesn't depend on $E_{\rm init}$ at all.  
All jets with a given initial opening angle
$\theta_{\rm jet}^{\rm init}$, with any value of the initial energy above the minimum possible,
suffer the same fractional energy loss if they traverse the same length of plasma.  In this
sense, jet energy loss is controlled by the initial opening angle of the jet and the trajectory
of the jet through the plasma, not by the initial energy of the jet.

In the right panel of Fig.~\ref{fig:IntroResults} we plot $E_{\rm jet}(x)/E_{\rm jet}^{\rm init}$.  Mirroring the behavior of $\theta_{\rm jet}$ above,
the jet energy decreases only slowly until $x/x_{\rm therm} \sim 0.5$.  
Indeed initially, $ \frac{dE_{\rm jet}}{dx} \sim x^2$ meaning that very little energy is lost.
In this regime we can expand and integrate (\ref{eq:energylossresult0}), obtaining
\begin{equation}
\label{eq:energylossresultSmallx}
\frac{\Delta E}{E_{\rm init}}  =  1 - \frac{4}{3\pi} \left( \frac{x}{x_{\rm therm}} \right)^3 - \frac{2}{5\pi}\left( \frac{x}{x_{\rm therm}} \right)^5
 - \frac{3}{14\pi}\left( \frac{x}{x_{\rm therm}} \right)^7 
- \ldots
\end{equation}
This expansion, to the order shown, deviates from the full result for $\Delta E$ obtained
by integrating (\ref{eq:energylossresult0}) by 0.2\% at $x/x_{\rm therm}=0.5$
and only by 2.5\% for $x/x_{\rm therm} = 0.75$, which
is to say by at most 2.5\% over distances for which $\Delta E/E_{\rm init}<0.22$.
However, as $x \to x_{\rm therm}$, $\frac{dE_{\rm jet}}{dx}$ diverges like $1/\sqrt{x_{\rm therm} - x}$,
meaning that the majority of the jet's energy is lost in the final stages of its trajectory.
In fact, more than three quarters of the initial energy of the jet is lost in the last quarter of $x_{\rm therm}$
and the last $17\%$ of the jet's energy is lost in the last $1\%$ of its trajectory.  
The rate of energy loss  increases until $x_{\rm therm} - x \sim 1/T$, at which point
the jet exits the SSR and the jet's instantaneous rate of energy loss becomes ill-defined.
Nevertheless, we see from (\ref{eq:energylossresult0}) and (\ref{eq:thermalizationdistance0}) that
for $x_{\rm therm} - x \gtrsim 1/T$, the jet still has a parametrically large amount of energy (compared to $T$),
meaning that the thermalization of the jet coincides with a dramatic burst of energy being transferred to the plasma.
This behavior, which was first suggested in \cite{Chesler:2008uy}, is reminiscent of a Bragg peak.

After setting up our calculation in Section~\ref{sec:elossdef} in the gauge theory,
and obtaining all the results above in Section~\ref{sec:GravitationalCalculation} via
holographic calculations, 
in Section~\ref{sec:concludingremarks} we shall close with some speculations about the qualitative lessons
for real jets, produced in heavy ion collisions, that may be learned from the results that we have summarized here.

We defer to an Appendix a discussion of why the ratio of the initial jet mass to the initial jet energy $M_{\rm init}/E_{\rm init}$ is {\it not}
a proxy for the initial jet angle $\theta_{\rm jet}^{\rm init}$ of the jets that we study, showing there that $M_{\rm init}/E_{\rm init}\propto (\theta_{\rm jet}^{\rm init})^{3/4}$
in the $\theta_{\rm jet}^{\rm init}\rightarrow 0$ limit.  This happens because these jets have stress energy flowing at angles
considerably larger than $\theta_{\rm jet}^{\rm init}$ when $\theta_{\rm jet}^{\rm init}$ is small, and these large-angle (maybe better
to say not-small-angle) tails of the jets
contribute to $M_{\rm init}/E_{\rm init}$ in a way that they do not
contribute to the jet opening angle itself.

\section{Defining the opening angle and energy loss rate in the field theory}
\label{sec:elossdef}

\begin{figure}
\begin{centering}
\includegraphics[width=10cm]{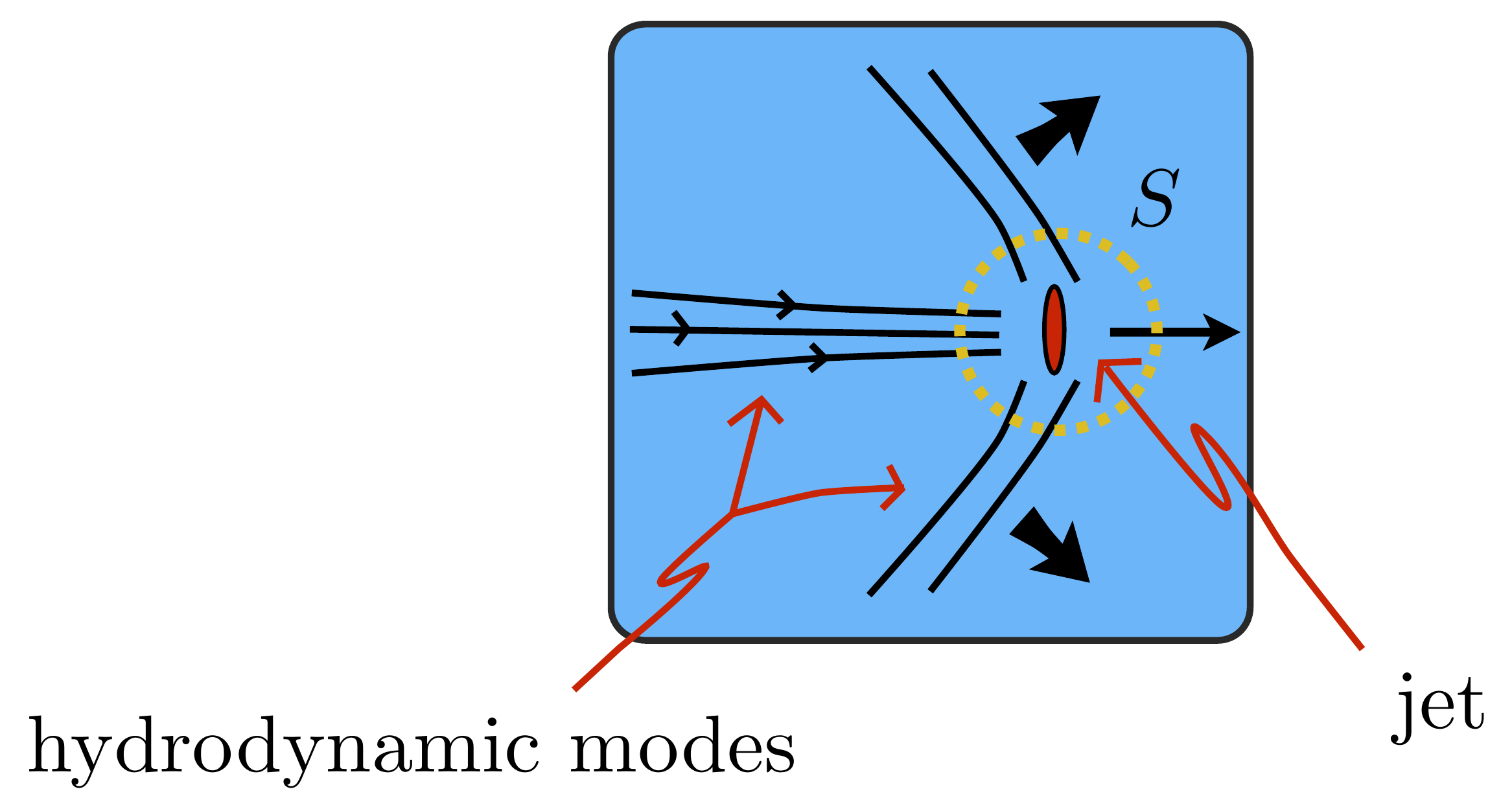}
\par\end{centering} 
\protect\caption{
\label{fig:jetcartoon} 
A cartoon showing a jet propagating through the strongly coupled fluid, losing
energy to hydrodynamic modes.  The cartoon is a snapshot at one
moment in time. The non-hydrodynamic components that constitute
the jet are found somewhere within the red oval.  (In QCD, the
non-hydrodynamic degrees of freedom inside the red oval  
would be the partons in
a shower. Here, we are describing the non-hydrodynamic stress energy
corresponding to an energetic light quark jet in strongly coupled ${\cal N}=4$ SYM theory.) 
The red oval encompassing
all  non-hydrodynamic stress energy is initially very small; as the jet propagates
from left to right, the red oval expands. As the jet thermalizes, after
traveling a distance $x_{\rm therm}\ggg 1/T$, the red oval has expanded
to a size of order $1/T$.
The dashed yellow sphere is
drawn at a radius such that the stress tensor is well described
by the constitutive relations of hydrodynamics at every point on it at all times.  It must have
a radius at least of order $1/T$, but can be chosen larger. 
}
\end{figure}


Consider a jet produced at $\bm x = \{x,\bm x_\perp\} = 0$ and propagating in the $x$-direction through an infinite and static plasma at temperature $T$.
Fig.~\ref{fig:jetcartoon} shows a far-zone cartoon of the jet at some fixed time.  The jet consists of a localized distribution of energy, within
the red oval in Fig.~\ref{fig:jetcartoon}, 
propagating in the $x$-direction whilst losing energy and momentum to the plasma.  The lost energy and momentum excite hydrodynamic 
modes in the plasma which in turn transport the lost energy and momentum away from the jet.  
In an infinite plasma, the jet propagates a distance $x_{\rm therm}$
before it has lost all of its energy and momentum and thermalized.
To have a well-defined jet and correspondingly, to crisply define the local energy loss
rate $dE_{\rm jet}(x)/dx$ and opening angle $\theta_{\rm jet}(x)$, it is necessary to 
consider jets whose $x_{\rm therm}$ is much greater than the plasma's characteristic microscopic scale, which at 
strong coupling is $1/T$.  Indeed, the scaling relation (\ref{eq:thermalizationdistance0}) means
this limit is realized by taking the initial jet energy 
$E_{\rm init}\ggg T\sqrt{\lambda}$, with $\lambda$ the 't~Hooft coupling of the gauge theory.
When $x_{\rm therm} \ggg 1/T$ both the rate of energy loss  and  the
opening angle must be slowly varying functions of $x$,  changing substantially only over scales of order $x_{\rm therm}$.  
In other words, over scales $\ell \ll x_{\rm therm}$, the evolution of the jet must be approximately 
steady-state with deviations from steady-state behavior suppressed by powers of $\ell/x_{\rm therm}$.  
Our calculations presented below show this remains true as long as $x$ lies in the SSR, Eq.~(\ref{SteadyStateRange}) 
above.  We shall obtain expressions for 
$dE_{\rm jet}(x)/dx$ and $\theta_{\rm jet}(x)$ within the domain
(\ref{SteadyStateRange}).

Before proceeding, it is worth noting that although we shall derive results valid throughout
the range (\ref{SteadyStateRange}), in drawing qualitative conclusions from our results
about the behavior of high energy jets in heavy ion collisions the full range is not relevant.
The highest energy jets that are observed in heavy ion collisions are, by definition, those
that have emerged from the droplet of plasma created in the collision long before
they travel a distance $x_{\rm therm}$.  

\begin{figure}
\begin{centering}
\includegraphics[width=11cm]{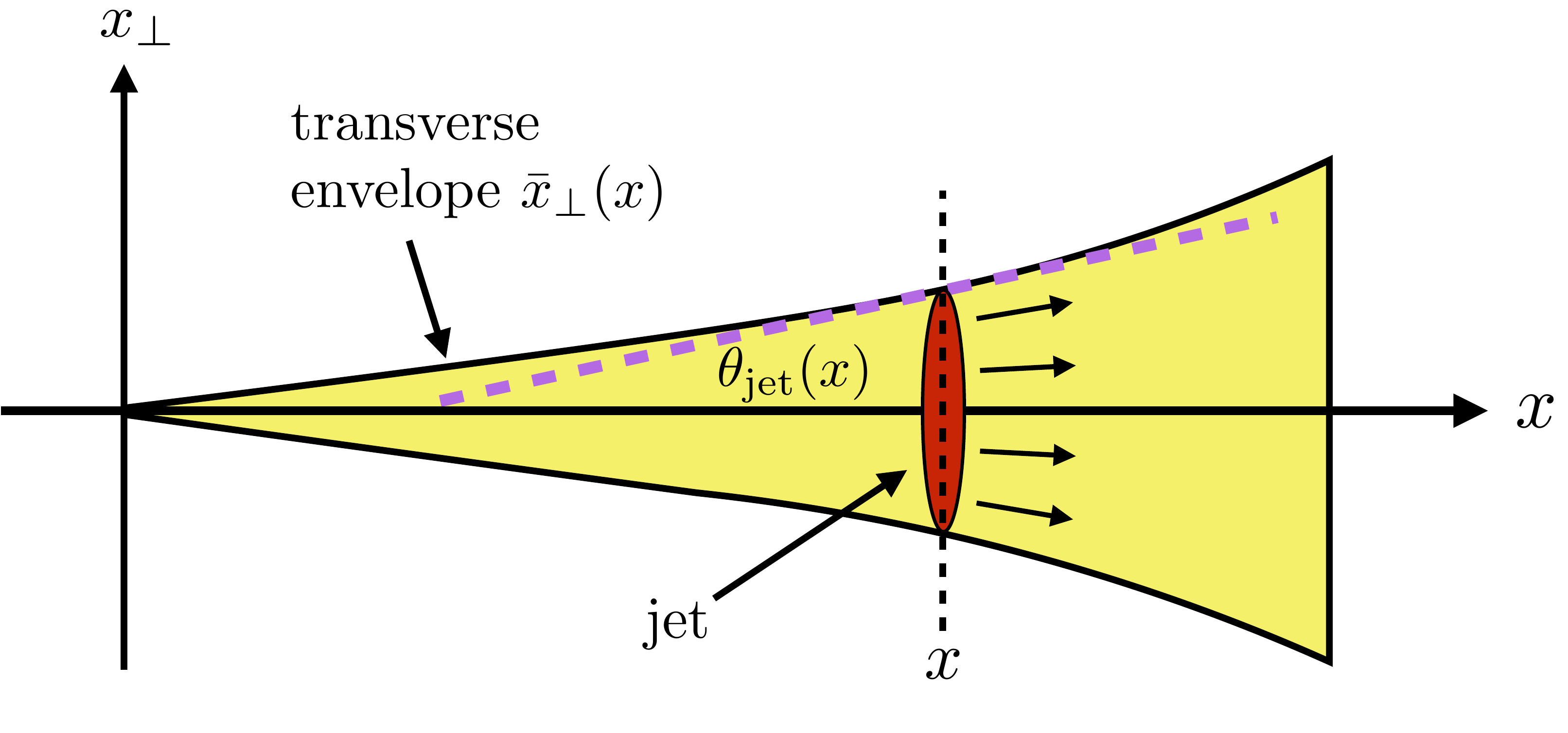}
\par\end{centering} 
\protect\caption{
\label{fig:jetcartoon0} 
A cartoon showing a jet, in red, at some fixed time.  The jet consists of a thin shell of energy that was produced at  
$x = x_\perp = 0$ and is
propagating in the $x$-direction, whilst expanding in the $\bm x_\perp$ directions.  
In the past and future, the jet was and will be
 located in the yellow shaded region bounded by the transverse envelope function $\bar x_\perp(x)$ defined in Eq.~(\ref{eq:envelope}).
The slope of the tangent line to the envelope function, shown as the dashed purple line, yields the instantaneous opening angle of the jet,
$\tan \theta_{\rm jet} = \frac{d \bar x_\perp}{dx}$.  In vacuum, $\bar x_\perp(x) \propto x$ and the opening angle is constant.
In plasma, however, the opening angle grows as a function of $x$ until $\bar x_\perp \sim 1/T$, at which point the jet thermalizes.
}
\end{figure}

Let us begin by defining the opening angle.  
Fig.~\ref{fig:jetcartoon0} shows a cartoon of the jet at some fixed time.  The jet consists of a thin shell of energy 
propagating in the $x$-direction whilst expanding in the transverse $\bm x_\perp$ directions. 
In the past and future the jet 
was and will be located in the yellow shaded region in the figure, which is bounded by some transverse envelope function $\bar x_\perp(x)$.
The slope of the tangent line to the envelope function, shown as the dashed purple line in the figure, 
defines the opening angle of the jet,
\begin{equation}
\label{eq:thetadef0}
\theta_{\rm jet}(x) \equiv \arctan \frac{d \bar x_{\perp}}{d x}\,.
\end{equation}
If the jet were in vacuum, $\theta_{\rm jet}(x)$  would be constant, meaning that the yellow shaded region
would be bounded by a straight line like the dashed purple line.  We can now be more
precise about what we mean by the phrase ``steady-state''.  In the steady-state region (\ref{SteadyStateRange}),
if we follow the jet only over some length scale $\ell \ll x_{\rm therm}$ its opening angle $\theta_{\rm jet}(x)$
does not change much, meaning that as it propagates over the distance $\ell$ its transverse extent
grows approximately linearly, as if in vacuum.  The effects of the presence of the plasma accumulate
over longer distances, as  $\theta_{\rm jet}(x)$ grows and the boundary of the yellow shaded region
curves outward.

To turn the picture in Fig.~\ref{fig:jetcartoon0} into a calculation of $\theta_{\rm jet}(x)$, we 
need a definition of the transverse envelope function $\bar x_\perp(x)$.
A simple definition of $\bar x_\perp$ comes from considering the 
flux of energy $\Phi$ through a surface at some constant $x$,
\begin{equation}
\label{eq:Phidef}
\Phi(\bm x) \equiv  \int dt \langle T^{0x}(t,\bm x) \rangle .
\end{equation}
The maximum of $\Phi$, which occurs at $x_\perp = 0$, must grow as the jet energy is taken
higher and higher and correspondingly, as $x_{\rm therm}$ is taken larger and larger.
We choose to define the transverse envelope $\bar x_\perp(x)$ by the half width at half maximum of $\Phi$,
\begin{equation}
\label{eq:envelope}
\Phi(x,\bar x_\perp) = \frac{1}{2} \Phi(x,x_\perp = 0).
\end{equation}
As we shall see below in Sec.~\ref{sec:openingangle}, in the SSR (\ref{SteadyStateRange}) we have
$\bar x_\perp \ll 1/T$, with $\bar x_\perp$ growing to order $1/T$ only when 
$x_{\rm therm} - x \sim 1/T$.

We now turn to the definition of the rate of energy loss.  As we have done for the jet opening
angle, we wish to define the rate of energy loss in the field theory, without introducing the
dual gravitational description.
A simple procedure to compute the energy loss rate is to surround the jet (the red oval in Fig.~\ref{fig:jetcartoon}) 
by a sphere $S$ 
and to compute the flux of energy through $S$,
\begin{equation}
\label{eq:elossdef1}
\frac{dE_{\rm jet}}{dt} = \int_S dS_i \langle T^{0 i} \rangle.
\end{equation}
The sphere $S$ is depicted as the yellow dashed circle in the cartoon shown in Fig.~\ref{fig:jetcartoon}.
It has a constant radius $R$ and is centered on the center of the jet at the time shown.
How big should $R$ be?  Since in the SSR the size of the jet is $\bar x_\perp \ll 1/T$, 
a natural choice is $R \gtrsim 1/T$.  
With this choice, the non-hydrodynamic physics of the jet itself, found within the red oval
in Fig.~\ref{fig:jetcartoon}, lies well inside the sphere $S$ at all times meaning that at all times
the
physics on the surface $S$ is described well by
hydrodynamics and the rate of energy loss (\ref{eq:elossdef1}) is simply the rate at which energy flows into hydrodynamic modes.  
However, it should be emphasized that in the SSR, the precise size of 
$S$ doesn't matter as long as $1/T \ll R \ll x_{\rm therm}$ since the jet evolution is steady-state over scales $\ll x_{\rm therm}$.

We shall compute the rate (\ref{eq:elossdef1}) at which energy flows into hydrodynamic modes 
in Fourier space.  Consider the Fourier transform
\begin{equation}
\langle \widetilde T^{\mu \nu}(\omega,\bm q) \rangle = \int dt \,d^3 x \, \langle T^{\mu \nu}(t,\bm x) \rangle e^{i \omega t} e^{-i \bm q \cdot \bm x},
\end{equation} 
and focus on slowly varying long wavelength modes with $x_{\rm therm} \omega$ and $x_{\rm therm} \bm q$ fixed in the large $x_{\rm therm}$ limit.
In this limit, the stress must take the form
\begin{equation}
\label{eq:longwavelengthexpansion}
\langle \widetilde T^{\mu \nu}(\omega,\bm q) \rangle = \widetilde T^{\mu \nu}_{\rm hydro}(\omega,\bm q) + \widetilde T^{\mu \nu}_{\rm jet}(\omega,\bm q),
\end{equation}
where
$\widetilde T^{\mu \nu}_{\rm hydro}$ satisfies the constitutive relations of hydrodynamics
and $\widetilde T^{\mu \nu}_{\rm jet}$ describes the nonhydrodynamic contribution to
the stress tensor.
Upon Fourier transforming 
back to real space, $T_{\rm jet}^{\mu \nu}(t,\bm x)$ 
must have a gradient expansion in terms of derivatives of delta functions
centered on the center-of-mass of the jet,
\begin{equation}
\bm x^{\rm CM}(t) \equiv \frac{\int d^3 \bm x \, \bm x \, T_{\rm jet}^{00}(t,\bm x)}
{\int d^3 \bm x \,  T_{\rm jet}^{00}(t,\bm x)}
\,.
\end{equation}
In other words, at low frequencies $\widetilde T^{\mu \nu}_{\rm jet}(\omega,\bm q)$ encodes the long wavelength limit of the 
localized stress near the jet.  Moreover, 
the total stress is conserved so
\begin{equation}
i q_\mu \widetilde T^{\mu \nu}_{\rm hydro} = - i q_\mu \widetilde T^{\mu \nu}_{\rm jet},  \ \ \  q^\mu = \{\omega,\bm q\},
\end{equation}
meaning $\widetilde T^{\mu \nu}_{\rm jet}$ sources the hydrodynamic flow.  At low frequencies, the Fourier transform of the 
jet energy, $\widetilde E_{\rm jet}(\omega)$, must be given by $E_{\rm jet}(\omega) = \lim_{\bm q \to 0} \widetilde T^{0 0}_{\rm jet}(\omega,\bm q)$.

What are the expected forms of $\widetilde T^{\mu \nu}_{\rm hydro}$ and $\widetilde T^{\mu \nu}_{\rm jet}$?
In the large $N_{\rm c}$ limit salient to holography,
the response of the plasma to the presence of the jet is $1/N_{\rm c}$ suppressed and the constitutive relations 
of linear hydrodynamics yield the leading order long wavelength expansion 
\begin{align}
\label{eq:hydroconstituative}
\widetilde T^{00}_{\rm hydro}(\omega,\bm q) &= \widetilde \varepsilon(\omega,\bm q),\nonumber\\
\widetilde T^{0i}_{\rm hydro}(\omega,\bm q) &= (\varepsilon_{\rm eq} + p_{\rm eq}) \widetilde {\bm u}(\omega,\bm q),\\
\widetilde T^{ij}_{\rm hydro}(\omega,\bm q) &= \widetilde p(\omega,\bm q) \delta^{ij},\nonumber
\end{align}
where $\widetilde \varepsilon$ and $\widetilde p$ are the proper energy and pressure, respectively, with $\varepsilon_{\rm eq}$ 
and $p_{\rm eq}$ their equilibrium values, and $\widetilde {\bm u}$ is the fluid velocity.
Conformal invariance in SYM requires $\widetilde T^{\mu \nu}_{\rm hydro}$ to be traceless, which means 
$\widetilde p = \widetilde \varepsilon/3$.  

Likewise, at small momentum the tensor structure of $\widetilde T^{\mu \nu}_{\rm jet}$ can only depend on the 
velocity of the jet's center of mass, $ V^\mu = \{1, {\bm V}\}$, and on the metric tensor $\eta^{\mu \nu}$.  
Demanding tracelessness then requires $\widetilde T^{\mu \nu}_{\rm jet} \propto V^\mu V^\nu + \frac{V_\alpha V^\alpha}{4} \eta^{\mu \nu}$.
However, for a jet composed of massless excitations, the velocity of
the jet's center of mass  
is related to the jet's  opening angle such that $V_\alpha V^\alpha\rightarrow 0$ for
jets whose $\theta^{\rm init}_{\rm jet}\rightarrow 0$,
meaning the center of mass of a well collimated jet
moves at approximately the speed of light.  More specifically,
we shall show in Appendix~\ref{app:JetMass} that 
\begin{equation}
V_\alpha V^\alpha = {\cal O}\left((\theta^{\rm init}_{\rm jet})^{3/2}\right)\ .
\label{eq:jetmassresult0}
\end{equation}
As 
$\theta_{\rm jet} \ll 1/(T x_{\rm therm})$, 
in the $x_{\rm therm} \gg 1/T$ limit we must simply have
\begin{equation}
\label{eq:jetstress}
\widetilde T^{\mu \nu}_{\rm jet} = \widetilde E_{\rm jet}(\omega - \bm V \cdot \bm q) V^\mu V^\nu.
\end{equation}
Note that $\widetilde T^{\mu \nu}_{\rm jet}$ only depends on the combination $\omega - \bm V \cdot \bm q$ because 
the jet center of mass trajectory, 
$x= \bm V t$, translates uniformly.

We therefore conclude that by computing the long wavelength limit of $\langle \widetilde T^{\mu \nu} \rangle$,
and then matching onto the expected forms (\ref{eq:longwavelengthexpansion}), (\ref{eq:hydroconstituative}) and (\ref{eq:jetstress}),
the low frequency limit of the jet energy $E_{\rm jet}(\omega)$ can be obtained.  More precisely, the real space function $\frac{dE_{\rm jet}}{dt}$ obtained 
in this manner is the derivative expansion of the energy loss in (\ref{eq:elossdef1}).  Nevertheless,
in the $x_{\rm therm} \gg 1/T$ limit, where the dynamics are approximately steady state, the 
two procedures for obtaining the energy loss rate must agree.

As a matter of convenience,  we will find it useful to extract $\widetilde E_{\rm jet}$  by projecting onto the traverse traceless mode of 
$\langle \widetilde T^{\mu \nu} \rangle$.
Let $\bm \epsilon_a$, $a = 1,2$ denote polarization vectors orthogonal to $\bm q$.
Define the transverse traceless mode of the stress, 
\begin{equation}
\label{eq:transversetracelessdef}
 {\mathcal T}_{ab}(\omega,\bm q) \equiv \left [\epsilon_a^i \epsilon_b^j - {\textstyle \frac{1}{2}} \delta_{ab}  \epsilon_c^i \epsilon_c^j \right]   
\langle \widetilde T_{i j}(\omega,\bm q) \rangle.
\end{equation}
The hydrodynamic stress in (\ref{eq:hydroconstituative}) lacks a transverse traceless mode, so  $ {\mathcal T}_{ab} = \mathcal T^{\rm jet}_{ab}$.
Substituting (\ref{eq:jetstress}) into (\ref{eq:transversetracelessdef}), we conclude that in the long wavelength limit,
the transverse traceless mode must read 
\begin{equation}
\label{eq:elossfromTTmode}
 {\mathcal T}_{ab}(\omega,\bm q) = \widetilde E_{\rm jet}(\omega - \bm V \cdot \bm q) \left [\epsilon_a^i \epsilon_b^j - {\textstyle \frac{1}{2}} \delta_{ab}  \epsilon_c^i \epsilon_c^j \right]  V^i V^j.  
\end{equation}

Our strategy for computing the opening angle and energy loss in holography is therefore to construct states where $x_{\rm therm} \ggg 1/T$.
For the opening angle we will compute the flux $\langle T^{0x} \rangle$ in the short wavelength limit near the jet
and construct the integrated flux $\Phi$ in (\ref{eq:Phidef}).  The opening angle is then given by (\ref{eq:thetadef0}) and (\ref{eq:envelope}).
In contrast, to compute the energy loss rate we will compute 
the long wavelength limit of the transverse traceless mode of the stress.  
The energy loss rate can then be extracted by matching the long wavelength result from holography 
onto the form of $\mathcal T_{ab}$  in (\ref{eq:elossfromTTmode}).
In this way, we will use a holographic calculation to determine both the opening angle and the rate of
energy loss entirely in terms of $\langle T^{\mu \nu}\rangle $.

%
%
%
%
%
%

\section{Gravitational calculation}
\label{sec:GravitationalCalculation}

According to gauge/gravity duality, the strongly coupled plasma of ${\cal N}=4$ SYM theory (infinite in extent,
static, in thermal equilibrium at a temperature $T$) is dual
to the 4+1-dimensional AdS-Schwarzschild black brane geometry~\cite{Witten:1998qj}, whose metric may be written
\begin{align}
\label{eq:metric}
&ds^2 = \frac{L^2}{u^2} \left [ -f dt^2 + d \bm x^2 + \frac{du^2}{f} \right ], \qquad
{\rm where} \qquad f \equiv 1 - \frac{u^4}{u_h^4},&
\end{align}
with $L$ the AdS radius.   The boundary of the geometry, which is where the dual 
field theory lives, is at AdS radial coordinate $u = 0$.
The geometry contains an event horizon at radial coordinate 
$u = u_h = 1/(\pi T)$ with $T$ the Hawking temperature of the black brane, which coincides with the 
temperature of the SYM plasma.

\begin{figure}
\begin{centering}
\includegraphics[width=12cm]{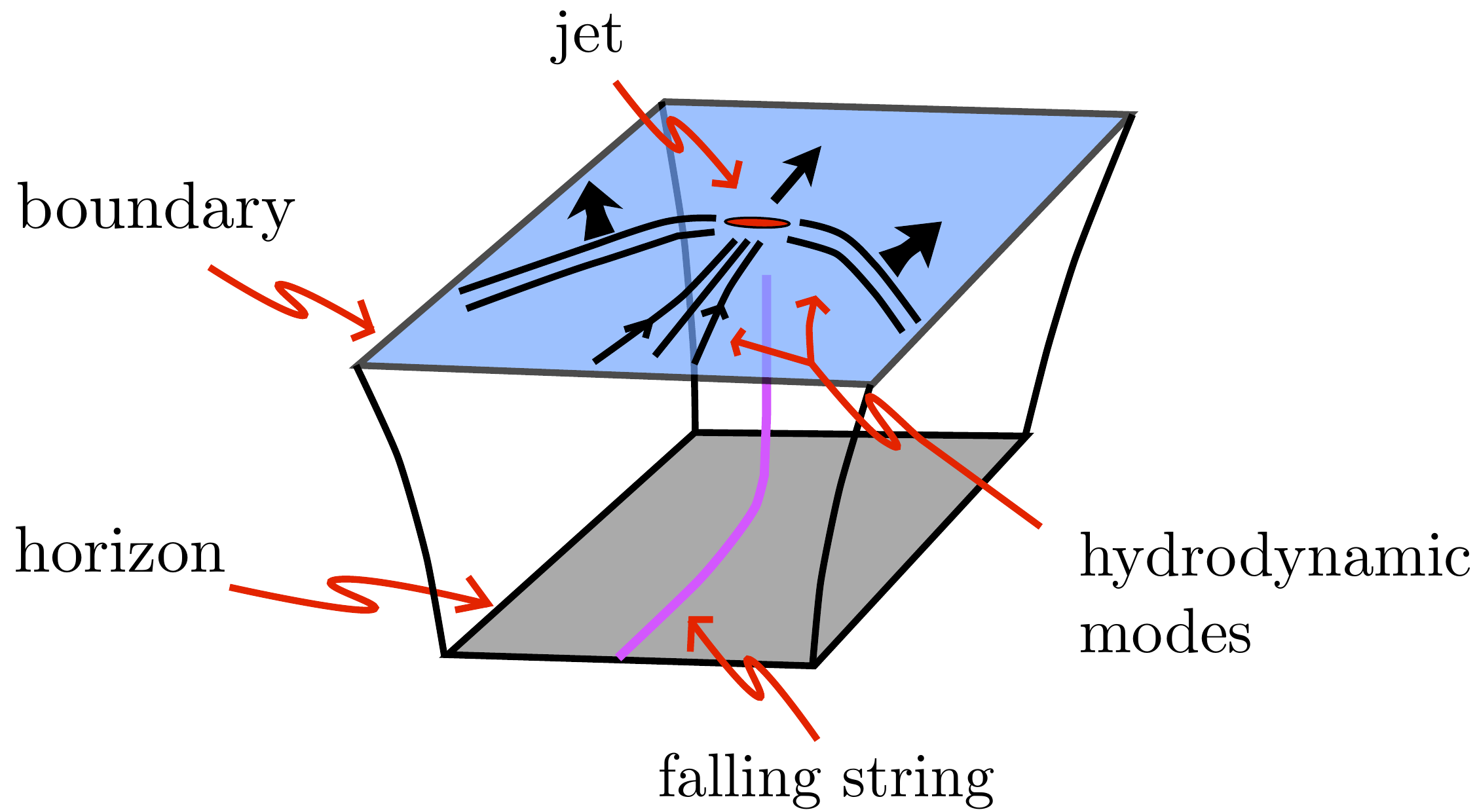}
\par\end{centering} 
\protect\caption{
\label{fig:bulkcartoon} 
A cartoon showing the gravitational description of a jet propagating through the strongly coupled fluid, losing
energy to hydrodynamic modes.  
The cartoon is a snapshot at one
moment in time.  The 
string is moving along the jet direction, as it falls.  The fact that the endpoint of the
string is falling as it moves corresponds, in the boundary theory, to the fact that
the red sphere encompassing all the nonhydrodynamic stress energy --- aka the jet --- expands as it moves.  
In the gravitational description, the falling string encodes the description of both the nonhydrodynamic stress
energy corresponding to the jet itself and the hydrodynamic stress corresponding
to its wake.
}
\end{figure}

Adding a massless quark jet to SYM plasma is equivalent to adding an open string 
to the black brane geometry \cite{Karch:2002sh}.  The string falls under the influence of gravity towards  
the black brane, with the approach to the event horizon encoding the thermalization of the jet in the field theory, see Fig.~\ref{fig:bulkcartoon}.
The presence of the string perturbs the metric $G_{MN}$,
\begin{equation}
\label{eq:linresponse}
G_{MN} = G_{MN}^{(0)} + \frac{L^2}{u^2} H_{MN},
\end{equation}
where $G_{MN}^{(0)}$ is the AdS-Schwarzschild metric (\ref{eq:metric}).  The metric perturbation $H_{MN}$
is governed by the linearized Einstein field equations,
\begin{equation}
\label{eq:linEinstein}
\mathcal L^{MN}_{AB} H_{MN} = 8 \pi G_{\rm Newton} J_{AB},
\end{equation}
where $\mathcal L^{MN}_{AB}$ is a linear differential operator 
(whose precise form follows from linearizing the Einstein equations about the AdS-Schwarzschild metric), 
$G_{\rm Newton}$ is the $5d$ gravitational constant and $J_{AB}$ is the $5d$ string stress tensor.

Let $\langle \Delta T^{\mu \nu} \rangle$ be the perturbation in the stress due to the presence of the jet,
\begin{equation}
\langle \Delta T^{\mu \nu} \rangle \equiv \langle  T^{\mu \nu} \rangle - \langle T^{\mu \nu}_{\rm eq} \rangle,
\end{equation}
where $ \langle T^{\mu \nu}_{\rm eq} \rangle$ is the equilibrium stress.
$\langle \Delta T^{\mu \nu} \rangle$ includes both hydrodynamic and nonhydrodynamic
contributions.
With the boundary conditions that 
$\lim_{u \to 0}H_{MN} = 0$, so the boundary geometry is simply that of Minkowski space, and in the gauge $H_{u M} = 0$, the linearized Einstein 
equations imply that $H_{\mu \nu}(t,\bm x,u) = u^4 H_{\mu \nu}^{(4)}(t,\bm x) + O(u^5)$.  In terms of the expansion coefficient $H_{\mu \nu}^{(4)}$,
the perturbation in the expectation value of the SYM stress tensor due to the presence of the jet reads \cite{deHaro:2000xn}
\begin{equation}
\label{eq:holorenorm0}
\langle \Delta T^{\mu \nu}(t,\bm x)\rangle = \frac{L^3}{4 \pi G_{\rm Newton}} H_{\mu \nu}^{(4)}(t,\bm x).
\end{equation}

Therefore, our strategy for computing $\langle T^{\mu \nu}\rangle$ will be  first to  construct string
states corresponding to the $x_{\rm therm} \ggg 1/T$ limit and then to solve the linearized Einstein equations
for the perturbation in the geometry due to the presence of the string.  We will then compute 
the jet's opening angle and energy loss rate from $\langle T^{\mu \nu} \rangle$.

\subsection{String dynamics}
\label{sec:stringdynamics}

The dynamics of the string are governed by the 
Nambu-Goto action
\begin{equation}
\label{eq:nambugoto}
S = -\frac{\sqrt{\lambda}}{2 \pi L^2} \int d \tau d \sigma \sqrt{-g}
\end{equation}
where $\lambda$ is the 't Hooft coupling, $\tau$ and $\sigma$
are worldsheet coordinates,
$g \equiv \det g_{ab}$, $g_{ab} \equiv \partial_a X \cdot \partial_b X$ is the string
worldsheet metric and $X^M = \{t(\tau,\sigma),x(\tau,\sigma),0,0,u(\tau,\sigma)\}$ are the string embedding functions.

Varying the action (\ref{eq:nambugoto}), we obtain the string equations of motion
\begin{equation}
\label{eq:stringeqm}
\partial_\tau \Pi^\tau_0 + \partial_\sigma \Pi^\sigma_0 = 0,
\end{equation}
and open string boundary conditions 
\begin{equation}
\label{eq:openbc}
\Pi^\sigma_M = 0 \ {\rm at \ string \ endpoints.}
\end{equation}
where $\Pi^a_M = \frac{\delta S}{\delta (\partial_a X^M)} $.  Explicitly,
\begin{subequations}
\label{eq:wscurrents}
\begin{align}
\label{eq:wsenergy}
\Pi^\tau_M &= - \frac{\sqrt{\lambda}}{2 \pi L^2} {  \frac{G_{MN}}{\sqrt{-g}}} \left [ (\dot X \cdot X') X'^N - (X')^2 \dot X^N \right ],
\\
\Pi^\sigma_M &= - \frac{\sqrt{\lambda}}{2 \pi L^2}  {  \frac{G_{MN}}{\sqrt{-g}}} \left [ ( \dot X \cdot  X') \dot X^N - (\dot X)^2  X'^N \right ].
\end{align}
\end{subequations}
Note that the energy of the string is
\begin{equation}
\label{eq:stringenergy}
E_{\rm string} = - \int d\sigma \Pi^{\tau}_0,
\end{equation}
so the string equations of motion (\ref{eq:stringeqm})
are simply the equations of energy conservation on the worldsheet,
with $\Pi^\sigma_0$ the energy flux.  

Following Refs.~\cite{Chesler:2008wd,Chesler:2008uy}, we model the creation of a massless quark and antiquark 
with a string 
created at a point asymptotically close to the boundary with initial condition $X^M |_{\tau = 0} = 0$
and with large momentum in the $\pm x$ directions.
The string subsequently expands into a finite size object in the $x{-}u$ plane as time progresses,
with endpoints moving apart in the $\pm x$ directions and falling towards the horizon.
The total distance the endpoints travel is simply the 
thermalization distance $x_{\rm therm}$~\cite{Chesler:2008uy,Gubser:2008as}.  The parametric relationship between the thermalization distance and the string 
energy is $x_{\rm therm} \sim E_{\rm string}^{1/3}$  \cite{Chesler:2008uy,Gubser:2008as}.

Strings whose $x_{\rm therm} \ggg 1/T$ have worldsheets which are approximately null 
\cite{Chesler:2008uy,Chesler:2014jva}.
Why?   When $x_{\rm therm} \to \infty$ the scaling $x_{\rm therm} \sim E_{\rm string}^{1/3}$ requires the string energy $E_{\rm string} \to \infty$.
Since strings have finite tension, the $E_{\rm string} \to \infty$ limit is generically realized by strings that expand at nearly the speed of light, 
meaning that the string profile must be approximately that of an expanding filament of
null dust.
Indeed, a null string profile $X_{\rm null}^M$ satisfies $g(X_{\rm null}) = 0$, and
from (\ref{eq:wsenergy}) has a  divergent energy density.
Our method of solving the string equations of motion closely mirrors that in our previous work~\cite{Chesler:2014jva}.
In particular, as we detail below, solving the string equations perturbatively about
a null configuration is tantamount to solving them using
geometric optics, with perturbations propagating on the
string worldsheet along null geodesics.

Since null strings satisfy $g(X_{\rm null}) = 0$, they minimize the Nambu-Goto action (\ref{eq:nambugoto}) 
and are exact, albeit singular, solutions to the string equations of motion (\ref{eq:stringeqm}).  
To obtain finite energy solutions to the equations of motion, we expand the string embedding functions about a null string solution
\begin{equation}
\label{eq:stringexpansion}
X^M = X^M_{\rm null} + \epsilon \,\delta X_{(1)}^M +  \epsilon^2 \delta X_{(2)}^M + \dots,
\end{equation}
where $\epsilon$ is a bookkeeping parameter (related to the string energy by $E_{\rm string} \sim 1/\sqrt{\epsilon}$) that
we shall treat as small for the purposes of organizing the non-linear corrections to the null string solution.
In what follows it is useful to choose 
worldsheet coordinates $\tau = t$ and $\sigma$ such that 
\begin{align}
\label{eq:stringexpansion2}
&\dot X_{\rm null} \cdot X'_{\rm null} = 0,&
\delta X_{(m)} = \{0,\delta x,0,0,0 \}.&
\end{align}
With this choice of worldsheet coordinates, the string endpoints cannot be at fixed $\sigma$.  Without loss of generality we shall focus on the right-moving
endpoint, whose location can be expanded in powers of $\epsilon$,
\begin{equation}
\sigma_{\rm endpoint} = \sigma_* + \epsilon \delta \sigma_{(1)} + \epsilon^2 \delta \sigma_{(2)} + \dots.
\end{equation}
The string equations of motion and boundary conditions
can then be solved perturbatively in powers of $\epsilon$.  The first step is constructing the null string $X_{\rm null}^M$.

The null string embedding functions can be written 
\begin{equation}
\label{eq:nullstringsolution}
X_{\rm null}^M = \{t,x_{\rm geo}(t,\sigma),0,0,u_{\rm geo}(t,\sigma)\},
\end{equation}
where for each $\sigma$, $x_{\rm geo}$ and $u_{\rm geo}$ satisfy the null geodesic 
equations which read
\begin{align}  
\label{eq:geo1}
&\frac{\partial x_{\rm geo}}{\partial t} = \frac{f}{\xi}, &
\frac{\partial u_{\rm geo}}{\partial t} = \frac{f\sqrt{\xi^2 -f }}{\xi},&
\end{align} 
where $\xi = \xi(\sigma)$.  The parameter $\xi$ determines the initial inclination of the geodesics in the $x{-}u$ plane
and, more fundamentally, specifies the conserved spatial momentum associated with the geodesics, $f(u)^{-1} \partial x_{\rm geo}/\partial t = \xi^{-1}$.
At leading order in $\epsilon$, the open string boundary conditions (\ref{eq:openbc}) are satisfied provided $\sigma_*$ is time-independent.

At early times $t \ll u_h$, when the string is close to the AdS boundary, the geodesics are given by 
\begin{align}
\label{eq:zeroTgeos}
&x_{\rm geo} = t \cos \sigma, & u_{\rm geo} = t \sin \sigma.&
\end{align}
Note that we have chosen constants 
of integration such that $x_{\rm geo}(t =0,\sigma) = u_{\rm geo}(t=0,\sigma) = 0$, meaning that the quark-antiquark pair is created at $x=0$.  
Hence, the worldsheet coordinate $\sigma$ is simply the initial angle of the geodesics in the $x{-}u$ plane.  Likewise,
\begin{equation}
\label{eq:xidef}
\xi(\sigma) = \sec(\sigma).
\end{equation}
Introducing the rescaled variables 
\begin{align}
\label{eq:hatteddefs}
&\hat u \equiv \frac{u_{\rm geo}}{u_h} \frac{1}{\sqrt{\tan \sigma}}\,,&
&\hat x \equiv   \frac{4 \sqrt{\pi}}{ \Gamma\left( {\textstyle \frac 14} \right)^2} \, \frac{x_{\rm geo}}{u_h}\,  \sqrt{\tan \sigma}\,, &
\end{align}
the solution to the geodesic equation (\ref{eq:geo1}) reads
\begin{equation}
\label{eq:geosol}
\hat x = F(\hat u),
\end{equation}
where 
\begin{equation}
\label{eq:Fdef}
F(\hat u) \equiv  1 -  \frac{4 \sqrt{\pi}}{ \Gamma\left(\textstyle \frac 14\right)^2}   \, \frac{ {_2F_1}\left({\textstyle \frac 14, \frac 12, \frac 54,-\frac{1}{\hat u^4}}\right)  }{\hat u}  \ .
\end{equation}

\begin{figure}
\begin{centering}
\includegraphics[width=12cm]{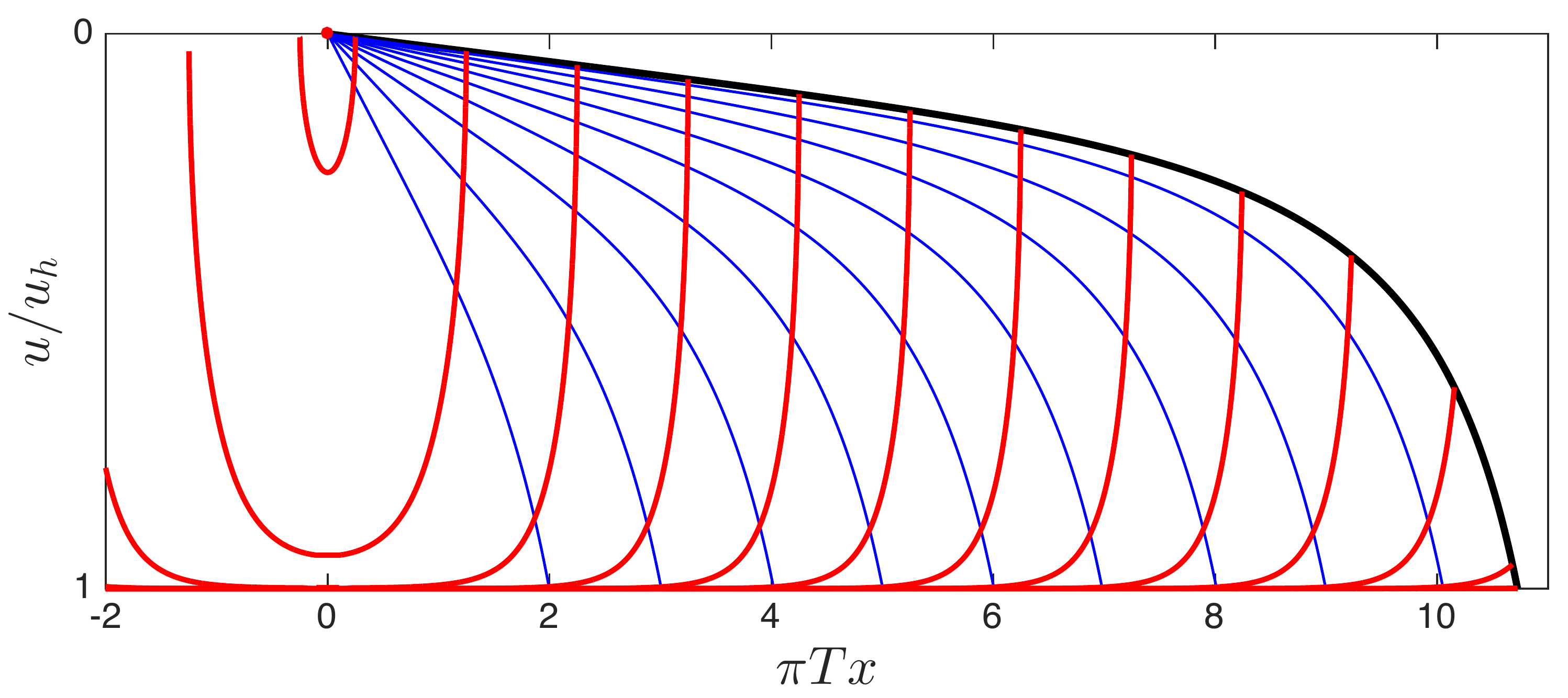}
\par\end{centering} 
\protect\caption{
\label{fig:NullString} 
A null string (red) shown at several different coordinate times $t$.
The string starts off at the point $x=0$ on the boundary and expands at the speed of light while falling towards the horizon.
The blue curves represent the null geodesics that each (red) bit of energy that makes up the string follows;
the black curve is the endpoint trajectory.  
Different blue curves are parametrized by different values of $\sigma$, where $\sigma$ is the initial angle
in the $(x,u)$ plane.  The endpoint is the trajectory with $\sigma=\sigma_*$; the figure is drawn with $\sigma_*=0.025$.
Due to the presence of the
horizon, which is to say due to the presence of the strongly coupled plasma, a blue (or black) trajectory with
a given $\sigma$ curves downward: its angle in the $(x,u)$ plane, which starts out equal to $\sigma$, steadily increases.
Clearly, geodesics with smaller  $\sigma$, i.e.~with smaller initial angle, propagate the farthest before reaching the horizon. It
is apparent in the figure that all the null geodesics strike the horizon at the same final angle. The steady-state region (SSR)
is the region where $x$ is not within of order $1/T$ of either $x=0$ or $x=x_{\rm therm}$.  In the limit in which $\sigma_*\to 0$ and
hence $x_{\rm therm}T\to\infty$, the string is in the SSR for almost its entire history.}
\label{fig:nullstring}
\end{figure}

In Fig.~\ref{fig:NullString} we plot a null string generated by a congruence of 
geodesics with $\sigma_* = 0.025$.  The string profile, denoted by the red curves, is shown at several values of coordinate time $t$.
The string starts off at a point on the boundary and expands at the speed of light while falling towards the horizon.
The blue curves represent the null geodesics followed by bits of the string 
and the black curve is the endpoint trajectory.
Every geodesic that makes up the string eventually falls into the horizon at some $x = x_{\rm stop}(\sigma)$.  Clearly, 
geodesics with smaller angle $\sigma$ go farther, with the endpoint geodesic going the farthest and reaching
the horizon after traveling 
a distance $x_{\rm therm}\equiv x_{\rm stop}(\sigma_*)$.  Indeed, by setting $u_{\rm geo} = u_h$, or equivalently $\hat u = \frac{1}{\sqrt{\tan \sigma}}$,
and expanding (\ref{eq:geosol}) about $\sigma = 0$, we obtain the stopping distance 
\begin{equation}
\label{eq:xstopsigma}
x_{\rm stop}(\sigma) = \frac{\Gamma({\textstyle \frac{1}{4}})^2}{4\pi T} \frac{1}{ \sqrt{\pi \sigma}}-\pi T + {\cal O}(\sigma^{1/2}).
\end{equation}
The leading order result for the  thermalization distance is then given simply by
\begin{equation}
\label{eq:xthermsigma}
x_{\rm therm} = x_{\rm stop}(\sigma_*) = \frac{\Gamma({\textstyle \frac{1}{4}})^2}{4\pi T} \frac{1}{ \sqrt{\pi \sigma_*}},
\end{equation}
a result obtained previously in Ref.~\cite{Chesler:2014jva}, and can be made arbitrarily large by taking the angle $\sigma_* \to 0$.
We therefore see that the $x_{\rm therm} \ggg 1/T$ limit is synonymous with the $\sigma_* \lll 1$ limit.
In what follows we shall use these two limits interchangeably.

If we denote the value of $\sigma$ that labels a 
geodesic that travels a total distance $x$ before plunging into the horizon by $\sigma_h(x)$,
from (\ref{eq:xstopsigma}) and (\ref{eq:xthermsigma}) we then see that
\begin{equation}
\label{eq:sigmaHresult}
\sigma_h(x) \equiv \sigma_* \left (\frac{x_{\rm therm}}{x} \right)^2 + {\cal O}\left(\sigma_*^{3/2}\right).
\end{equation}
Recall that we can isolate the SSR of Eq.~(\ref{SteadyStateRange}) by taking the
$x_{\rm therm} T \to \infty$ limit, which is to say the $\sigma_*\to 0$ limit, while keeping
$x/x_{\rm therm}$ fixed at any value between 0 and 1.  
Noting that above-horizon geodesics are those with $\sigma < \sigma_h(x)$, 
we see from (\ref{eq:sigmaHresult})
that in this limit, all above-horizon geodesics have $\sigma = {\cal O}(\sigma_*)$.


What is the above-horizon shape of the string shown in Fig.~\ref{fig:NullString}?  The solution to the geodesic 
equation (\ref{eq:geosol}) can also be written
\begin{equation}
x_{\rm geo} = \xi t + x_0(u_{\rm geo}),
\end{equation}
where $x_0$ satisfies 
\begin{equation}
\label{eq:stringprofile}
\frac{\partial x_0}{\partial u_{\rm geo}} = - \frac{\sqrt{\xi^2 - f}}{f}.
\end{equation}  
Consider the $\sigma \to 0$ limit of (\ref{eq:stringprofile}). 
When $u \sim \sqrt{\sigma} u_h$ we have
$\frac{\sqrt{\xi^2 - f}}{f} \approx \sqrt{\xi^2 - f}$ and
the solution to (\ref{eq:stringprofile})
can be expressed in terms of hypergeometric functions.  When $u \gg \sqrt{\sigma} u_h$ we can approximate $\xi \approx 1$ and 
$\frac{\sqrt{\xi^2 - f}}{f} \approx \frac{u^2}{u_h^2 f}$, so 
the solution to (\ref{eq:stringprofile}) reads 
\begin{equation}
\label{eq:trailing}
x_0(u_{\rm geo})= \frac{u_h}{2} \left ( \arctan \frac{u_{\rm geo}}{u_h} - {\rm arctanh}\, \frac{u_{\rm geo}}{u_h} \right ),
\end{equation}
which is simply the trailing string profile of Refs.~\cite{Herzog:2006gh,Gubser:2006bz}.  Therefore, when
the string is in the SSR,
everywhere except very close to its endpoint,
its profile illustrated in Fig.~\ref{fig:NullString},
is simply the trailing string profile, suitably truncated to reflect the falling endpoint, and translating at the speed of light.

We turn now to the first order correction $\delta x_{(1)}$ defined in (\ref{eq:stringexpansion2}). 
At leading order in the bookkeeping 
parameter $\epsilon$, the worldsheet energy density $\Pi^\tau_0$ and flux $\Pi^\sigma_0$ read
\begin{align}
\label{eq:Pi00}
&\Pi^\tau_{0} = - \frac{\sqrt{\lambda}}{2\pi } \frac{ \xi \partial_\sigma u_{\rm geo}}{u_{\rm geo}^2} \sqrt{ \frac{- \xi}{2 \epsilon f \partial_t \delta x_{(1)}}} + {\cal O}(\sqrt{\epsilon}), &
\Pi^\sigma_0 = {\cal O}(\sqrt{\epsilon}).&
\end{align}
Hence, at leading order in $\epsilon$ the string equations of motion (\ref{eq:stringeqm}) simply read $\partial_t \Pi^\tau_0 = 0$, 
meaning $\Pi_0^\tau(t,\sigma) = \Pi^\tau_0(\sigma)$.
In other words, energy is  transported on the congruence of geodesics which make up the null string.
This is a consequence of causality: in the limit where the string expands asymptotically close to the speed of 
light, different points on the string are causally disconnected from each other and the energy of each bit of string must be transported along the 
null rays that describe how the string expands.

Because $\Pi^\tau_0$ is time independent, it is only necessary to compute it at one time,
which for convenience we take to be an early time when the entire string is near the AdS boundary.
Near the AdS boundary, where $f = 1$ and the geodesics are given by (\ref{eq:zeroTgeos}), the string equation of motion $\partial_t \Pi^\tau_0 = 0$ 
leads to the equation of motion for $\delta x_{(1)}$
\begin{equation}
\partial_t^2 \delta x_{(1)} + \frac{2}{t} \partial_t \delta x_{(1)} = 0,
\end{equation}
which has the solution 
\begin{equation}
\label{eq:stringsol2}
\delta x_{(1)}(t,\sigma) = \phi(\sigma) + \frac{1}{t} \psi(\sigma),
\end{equation}
for arbitrary functions $\phi(\sigma)$ and $\psi(\sigma)$.  
Upon substituting the string solution (\ref{eq:stringsol2})
and the geodesic congruence (\ref{eq:zeroTgeos}) and (\ref{eq:xidef}) into (\ref{eq:Pi00}), 
we obtain
\begin{equation}
\label{eq:pi2}
\Pi^\tau_0(\sigma) = -\frac{\sqrt{\lambda}}{2 \pi} \csc^2 \sigma \sqrt{\frac{\csc 2 \sigma \sin \sigma}{ \epsilon \psi(\sigma)}} + {\cal O}(\sqrt{\epsilon}).
\end{equation}

We now turn to enforcing the open string boundary conditions.
At first order in $\epsilon$ and near the AdS boundary, where $f = 1$, the open string boundary conditions (\ref{eq:openbc})
require 
\begin{align}
&\psi(\sigma_*) = 0,
&\delta \sigma_{(1)} = \frac{2 t \, \phi'(\sigma_*) + \psi'(\sigma_*)}{2 t^2} \cos \sigma_*.
\end{align}
How does $\psi(\sigma)$ vanish as $\sigma \to \sigma_*$?  We see from
the expression (\ref{eq:pi2}) for the density of energy along the string
per unit $\sigma$ that finiteness of the string energy 
requires $\psi(\sigma) \sim (\sigma - \sigma_*)^p$ with $p < 2$.  Moreover, 
note that in general $\delta \sigma_{(n)}$ contains a term proportional to $\partial_\sigma^n \psi|_{\sigma = \sigma_*}$.
Finiteness of $\delta \sigma_{(n)}$ and finiteness of the total string energy therefore require 
$\psi$ to have a near-endpoint asymptotic expansion of the form
\begin{equation}
\label{eq:psiexpansion}
\psi(\sigma) = \sum_{k = 1}^{\infty} \psi_k (\sigma - \sigma_*)^k.
\end{equation}

In the SSR, where geodesics with $\sigma = {\cal O}(\sigma_*)$ make up the string, 
we may employ the expansion (\ref{eq:psiexpansion}), keeping only the leading term.
Expanding (\ref{eq:pi2}), we then obtain the above-horizon world sheet energy density
\begin{equation}
\label{eq:Pi00UV}
\Pi_0^\tau(\sigma) = -\frac{\sqrt{\lambda}}{2 \pi} \frac{1}{\sqrt{2 \epsilon \,\psi_1}} \frac{1}{\sigma^2 \sqrt{\sigma - \sigma_*}}\left [1 - {\cal O}(\sigma-\sigma_* ) \right ].
\end{equation}
We therefore see that up to an overall constant specified by $\psi_1$, the world sheet energy density is uniquely determined in the SSR. 
Simply put, the large $\sigma$ information contained in $\psi$ and $\Pi^\tau_0$ falls into the horizon promptly after the string is created,
before the string enters the SSR.

At leading order in $\epsilon$, the string stress tensor which sources the linearized Einstein field equations  is given by 
\begin{equation}
\label{eq:stringstress}
J^{MN} = \int d\sigma J^{M N}_{\rm particle}(\sigma)
\end{equation}
where 
\begin{equation}
\label{eq:particlestress}
J^{MN}_{\rm particle} = 
 \frac{\Pi^{\tau}_{0}}{G_{00}} \frac{dX_{\rm geo}^M}{dt} \frac{dX_{\rm geo}^N}{dt}  \frac{1}{\sqrt{-G}}   \delta^3(\bm x - \bm x_{\rm geo}) \delta(u - u_{\rm geo}),
\end{equation}
is just the stress for a single null particle moving on a geodesic labeled by $\sigma$ with energy 
$\varepsilon_{\rm particle} = - \int d^3 x\, du\, \sqrt{-G} J^{0}_{\ 0} = - \Pi^\tau_0$.  

\subsection{The evolution of the jet envelope and opening angle}
\label{sec:openingangle}

We wish to solve the gravitational bulk to boundary problem in the SSR.
This requires focusing on  $x_{\rm therm} \ggg 1/T$, or equivalently, when $\sigma_* \lll 1$.
Before proceeding with the gravitational bulk to boundary problem, 
let us analyze the congruence of geodesics which make up the above-horizon segment of string.
As discussed above, these geodesics have $\sigma_* \leq \sigma < \sigma_h(x)$ with 
$\sigma_h(x)$ given in (\ref{eq:sigmaHresult}).  However, 
in the SSR it turns out that only geodesics with $\sigma$ parametrically close to 
$\sigma_h$ are close to the horizon.  Geodesics with $\sigma$ not close to $\sigma_h$ are close to the boundary with
\begin{equation}
\label{eq:SteadyStateRangeBulk}
u_{\rm geo} \sim \sqrt{\sigma} u_h \sim \sqrt{\sigma_*} u_h.
\end{equation}
To verify these statements we employ the rescaled variables $\hat x$ and $\hat u$ defined in Eq.~(\ref{eq:hatteddefs})
and the geodesic solution (\ref{eq:geosol}).   
From (\ref{eq:hatteddefs}) and (\ref{eq:sigmaHresult}) we have 
\begin{equation}
\label{eq:xhatsigmah}
\hat x = \sqrt{\frac{\sigma}{\sigma_h(x_{\rm geo})}} + {\cal O}(\sqrt{\sigma_*}).
\end{equation} 
Moreover, as $\hat x \to 1$, Eq.~(\ref{eq:geosol}) implies that $\hat u$ diverges like
\begin{equation}
\label{eq:uhatdivergence}
\hat u \sim \frac{1}{1 - \hat x}.
\end{equation} 
It therefore follows that $\hat u \sim \sigma^0_*$ and, correspondingly, $u_{\rm geo} \sim \sqrt{\sigma_*}u_h$ until $\sigma$ becomes 
parametrically close to $\sigma_h$.  In particular, 
upon substituting (\ref{eq:xhatsigmah}) into (\ref{eq:uhatdivergence}) we see that
only geodesics with  $\frac{\sigma_h - \sigma}{\sigma_h} \sim \sqrt{\sigma}$
have $\hat u \sim 1/\sqrt{\sigma}$ and, correspondingly, $u_{\rm geo} \sim u_h$.  
Since the string's energy density $\Pi^\tau_0$ is simply transported along the geodesics which make up the null string, 
and since $\Pi^\tau_0$ is greatest near the endpoint,
we  conclude that in the $\sigma_* \to 0$ limit a parametrically large fraction of the 
string's above-horizon energy is located asymptotically close to the AdS boundary.  

Via the gravitational bulk to boundary problem, the near-endpoint gravitational field of the string 
induces a highly peaked and localized 
stress on the boundary.  The localized stress induced by the near-endpoint segment 
of the string must determine the jet opening angle and energy.
Since the vast majority of the string's above-horizon energy is located asymptotically close to the AdS boundary,
the scale of localization of the jet must be $\ll 1/T$.  It follows that in the region of space where the jet is localized, 
one can employ zero temperature Green functions (which can be computed analytically) to solve the linearized Einstein equations
and compute the near-jet boundary stress tensor.  
That is, close to the AdS boundary the geometry is approximately that of AdS$_5$ and 
the gravitational bulk to boundary propagators take their zero temperature 
form.
However, the source for the linearized Einstein equations --- the string stress tensor --- incorporates 
finite temperature effects that cannot be neglected.  In particular, the congruence of geodesics that make up
the near-endpoint segment of the string are pulled towards the horizon by the gravitational field of the black hole
and this effect accumulates as the string  traverses the SSR and cannot be neglected.
As we shall see below, in the boundary theory the bending and falling of geodesics towards the horizon
encodes the broadening of the jet opening angle.

In characterizing the bending of a geodesic towards the horizon it is useful to define the angle 
\begin{equation}
\sigma_{\rm eff}(\sigma,x)\equiv \arctan \frac{du_{\rm geo}}{dx_{\rm geo}}\bigg |_{x_{\rm geo} = x}\, ,
\label{eq:sigmaeff}
\end{equation}
 which is simply the angle a geodesic labeled by $\sigma$ makes with the boundary at point $x$.
At small $x$ we have $\sigma_{\rm eff} = \sigma$.  
If the geodesic were propagating in vacuum, we would have $\sigma_{\rm eff} = \sigma$ for all time.
However, as the geodesic propagates through the plasma it curves downward toward the horizon, meaning that
$\sigma_{\rm eff}$ increases.  
Using the geodesic solution (\ref{eq:geosol}) and the definition (\ref{eq:sigmaeff}), it can be shown that  
\begin{equation}
\label{eq:sigmaeff1}
\tan \sigma_{\rm eff}(\sigma,x) = \tan \sigma \sqrt{1 + \hat u(\sigma,x)^4},
\end{equation}
with 
\begin{equation}
\label{eq:invgeosol}
\hat u(\sigma,x) = F^{-1} (\hat x(\sigma,x)),
\end{equation}
where $F^{-1}$ is the inverse of the function $F$ defined in (\ref{eq:Fdef}).  
The expression (\ref{eq:sigmaeff1})
describes how $\sigma_{\rm eff}$ increases and
the geodesic curves downward as it propagates.
One can show directly from the geodesic solution (\ref{eq:geosol})
that at $x_{\rm geo}=x_{\rm stop}$, where $u_{\rm geo}=u_h$, the geodesic has $du_{\rm geo}/dx_{\rm geo}=\sec \sigma$,
meaning that at the point where the geodesic strikes the horizon
$\sigma_{\rm eff}$ has increased to a final value 
$\sigma_{\rm eff}(\sigma,x_{\rm stop}) = \arctan \sec \sigma$, which is to say $ \sigma_{\rm eff}(\sigma,x_{\rm stop}) \approx \frac{\pi}{4}$
since $\sigma$ is small.  This explains why all the blue curves in the SSR in
Fig.~\ref{fig:nullstring} strike the horizon at the same angle.  Although $\sigma_{\rm eff}$ rapidly increases
to its final value as the geodesic approaches the horizon, by differentiating (\ref{eq:sigmaeff1}) we see that
at earlier times when the geodesic is 
in the near-boundary domain (\ref{eq:SteadyStateRangeBulk}),
$\sigma_{\rm eff}$ is slowly varying with $\frac{\partial \sigma_{\rm eff}}{\partial x} \sim \sigma \frac{\partial \hat x}{\partial x} \sim \frac{\sigma^{3/2}}{u_h}
\sim \frac{\sigma}{x_{\rm therm}}$.
Because of this, locally near any point $x$ we may approximate the near-boundary geodesics with their tangent lines
\begin{align}
\label{eq:nearboundarygeos}
&x_{\rm geo} \approx (t - \Delta t) \cos \sigma_{\rm eff} + \Delta x,&
&u_{\rm geo} \approx (t - \Delta t) \sin \sigma_{\rm eff},
\end{align}
where $\Delta x$, $\Delta t$ and $\sigma_{\rm eff}$ all depend on $\sigma$ and $x$.
In other words, locally the geodesics take their zero temperature form (\ref{eq:zeroTgeos}) with accumulative finite temperature effects 
encoded in $\sigma_{\rm eff}$ and $\Delta x$ and $\Delta t$.  

A consequence of the above analysis is that the near-boundary segment 
of string, where nearly all the above-horizon string energy is located, is 
just a sum of null point particles moving on trajectories which locally 
take the zero temperature form (\ref{eq:nearboundarygeos}) 
and which have energy $\varepsilon_{\rm particle}(\sigma) = - \Pi^\tau_0(\sigma).$
As we noted above, in this near-boundary region it is appropriate to use the zero temperature
gravitational bulk to boundary propagators.  The boundary stress tensor induced by a single null 
particle of energy $\varepsilon_{\rm particle}$
falling in the zero temperature AdS$_5$ geometry along a geodesic (\ref{eq:nearboundarygeos})
was computed in Ref.~\cite{Hatta:2010dz}.  Their result reads
\begin{equation}
\label{eq:particlestress2}
\langle T^{\mu \nu}_{\rm particle} \rangle = \frac{\varepsilon_{\rm particle}}{4 \pi |\Delta \bm x|^2} \frac{\sin^4 \sigma_{\rm eff} }{(1 - \Delta \hat x \cdot \bm v)^3} 
\frac{ \Delta x^\mu  \Delta x^\nu}{|\Delta \bm x|^2} \delta(t -\Delta t- |\Delta \bm x|),
\end{equation}
where $\Delta x^\mu \equiv \{t - \Delta t, \Delta \bm x\}$,
$\Delta \bm x \equiv  \{x - \Delta x,\bm x_\perp\}$, 
$\Delta \hat x \equiv \Delta \bm x/ |\Delta \bm x|$,
and $v \equiv \frac{d x_{\rm geo}}{dt}  = \cos \sigma_{\rm eff}$.  By linearity, the full expression for the stress induced by the near-endpoint segment of the string reads
\begin{equation}
\label{eq:DeltaStress}
\langle T^{\mu \nu}_{\rm near-jet}\rangle = \int_{\sigma_*}^{\sigma_h} d\sigma \langle T^{\mu \nu}_{\rm particle} \rangle.
\end{equation}
The integrated flux $\Phi$ through a surface of constant $x$, Eq.~(\ref{eq:Phidef}), therefore reads
\begin{equation}
\label{eq:holoflux}
\Phi = \frac{1}{4 \pi} \int d \sigma 
\frac{-\Pi^\tau_0 \sin^4 \sigma_{\rm eff} }{(1 - \Delta \hat x \cdot \bm v)^3} 
\frac{x - \Delta x}{|\Delta \bm x|^3}.
\end{equation}
As we shall see below, in the $\sigma_* \to 0$ limit $\Phi$ has a parametrically high amplitude $\Phi(x,x_\perp = 0) \sim \sigma_*^{-5/2}$
with a parametrically small width $\bar x_\perp \sim \sqrt{\sigma_*}$.  
This justifies neglecting all other contributions to the flux except those coming from the near-endpoint segment 
of the string.

To proceed further it is useful to expand (\ref{eq:holoflux}) in powers of $\sigma \sim \sigma_*$.
To do so we assume $x/x_{\rm therm} = {\cal O}(1)$ so $x \sim u_h/\sqrt{\sigma_*}$ and that $x_\perp \sim \sqrt{\sigma_*} u_h$.
At small $\sigma$ and in the near-boundary region (\ref{eq:SteadyStateRangeBulk}), where $\hat u = {\cal O}(\sigma^0)$,
we may approximate (\ref{eq:sigmaeff1}) as 
\begin{equation}
\label{eq:sigmaeff2}
\sigma_{\rm eff}(\sigma,x) =  \sigma \sqrt{1 + \hat u(\sigma,x)^4}.
\end{equation}
Likewise, in the small $\sigma_*$ limit we may use (\ref{eq:xhatsigmah}) to approximate
\begin{equation}
\label{eq:hatuexpansion}
 \hat u(\sigma,x) = F^{-1} \left( \textstyle \sqrt{\frac{\sigma}{\sigma_h(x)}}\right).
 \end{equation}
Turning next to $\Delta x$ and $\Delta t$, it follows from the geodesic equation (\ref{eq:geo1})
that $dx_{\rm geo}/dt \approx 1$ in the near-boundary region  (\ref{eq:SteadyStateRangeBulk}).
This means $\Delta x = \Delta t$.  Likewise, it follows from the solution 
(\ref{eq:geosol}) to the geodesic equation that
in the small $\sigma$ limit we have
\begin{equation}
\label{eq:DeltatDeltax}
\Delta t(\sigma,x) = \Delta x(x,\sigma) = \frac{u_h}{\sqrt{\sigma}} \left [ { \frac{ \Gamma( \frac 14)^2}{4 \sqrt{\pi}} \sqrt{\frac{\sigma}{\sigma_h(x)}}} - \frac{\hat u(\sigma,x)}{\sqrt{1 + \hat u(\sigma,x)^4}} \right ].
\end{equation}
Upon substituting Eqs.~(\ref{eq:sigmaeff2}) and (\ref{eq:DeltatDeltax}) and the world sheet energy density (\ref{eq:Pi00UV})
into (\ref{eq:holoflux}) and expanding in powers of $\sigma_*$, we secure the leading order result 
\begin{equation}
\label{eq:holoflux2}
\Phi(x,x_\perp) =  \sqrt{ \frac{ {\lambda}}{2 \pi^4 \epsilon\, \psi_1}} 
\int_{\sigma_*}^{\sigma_h(x)} d\sigma \frac{\hat u(\sigma,x)^4}{u_h^2 \left [  \sigma \hat u(\sigma,x)^2 +   x_{\perp}^2/u_h^2 \right ]^3} \frac{1}{ \sqrt{\sigma - \sigma_*}},
\end{equation}
again with $\hat u(\sigma,x)$ given by (\ref{eq:hatuexpansion}).


Before proceeding further, several comments are in order:
\begin{enumerate}

\item We see from (\ref{eq:sigmaHresult}) and (\ref{eq:hatuexpansion}) that 
the integral in (\ref{eq:holoflux2}) only depends on $x$ via the ratio $x/x_{\rm therm}$.  This means that upon making the rescalings 
$\sigma \to \sigma/\sigma_*$ and $x_\perp \to x_{\perp}/\sqrt{\sigma_*}$ and keeping $x/x_{\rm therm}$ fixed, 
the integral in (\ref{eq:holoflux2}) is given by $\sigma_*^{-5/2}$ multiplying an expression that has a finite $\sigma_* \to 0$ limit.  
Hence, as advertised above, as $\sigma_*\to 0$ we find that $\Phi(x,x_\perp)$ has a peak value at $x_\perp=0$
that grows like $\sigma_*^{-5/2}$ and  
has a parametrically small 
width in $x_\perp$ that scales like $\sigma_*^{1/2}$.

\item The integrand in (\ref{eq:holoflux2}) vanishes as $\sigma \to \sigma_h$ where, according to Eqs.~(\ref{eq:xhatsigmah}) and (\ref{eq:uhatdivergence}), $\hat u$ grows unboundedly large.
More precisely, the integrand at $\sigma = \sigma_h$ is suppressed in value 
since $\hat u = 1/\sqrt{\tan \sigma} \sim 1/\sqrt{\sigma_*}$ at the horizon and we took the $\sigma_* \to 0$ limit to derive (\ref{eq:holoflux2}).
This means that contributions to $\Phi$ from the near-horizon segment of the string are suppressed in the $\sigma_*\to 0$ limit 
relative to those from the near-endpoint segment of the string.

\begin{figure}
\begin{centering}
\includegraphics[width=11cm]{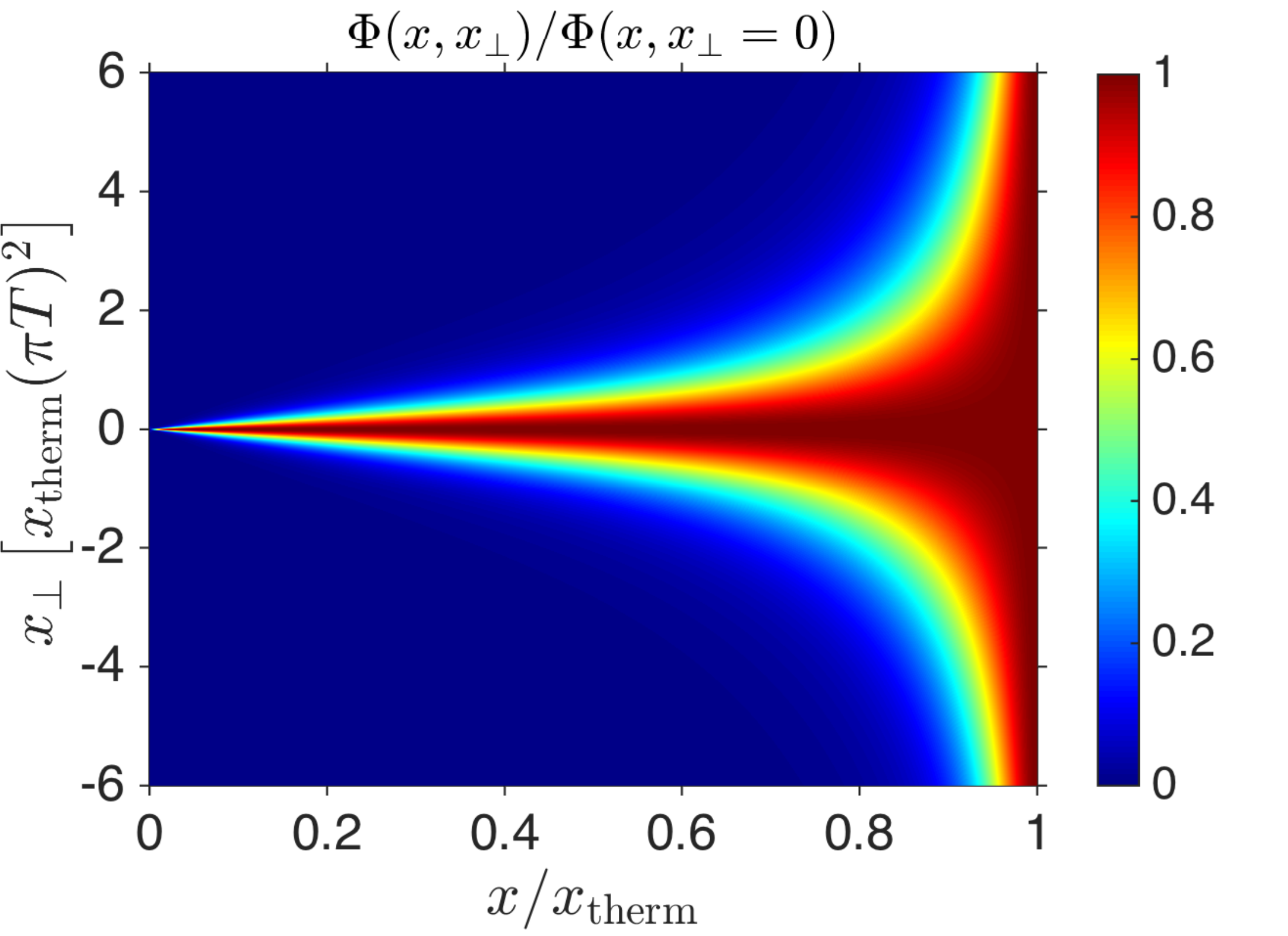}
\par\end{centering} 
\protect\caption{
\label{fig:NormalizedPhi} 
The color at a point in this plot  depicts $\Phi(x,x_\perp)$, the integrated flux of energy in the $x$-direction
that passes through a point on the
planar surface at a given value of $x$ that is a 
given transverse distance $x_\perp$ from the center of the jet.
At each value of $x$, we normalize $\Phi(x,x_\perp)$ relative to its maximum
value at that $x$, namely $\Phi(x,0)$.  We have normalized $x_\perp$ by a factor of $x_{\rm therm}$
in order to obtain a figure that is unchanging as the $\sigma_*\to 0$ limit is taken.   This figure should
be compared to the cartoon in Fig.~\ref{fig:jetcartoon0}.   We shall take as our quantitative definition of the yellow region
in that cartoon the region of the present figure in which $\Phi(x,x_\perp)$ is more than half its maximum
value $\Phi(x,0)$; we could equally well have chosen a definition in which ``half'' was replaced, for example, by 10\%.
}
\end{figure}

\item 
We do not know how to evaluate the integral in (\ref{eq:holoflux2})
analytically.  Nevertheless, by performing the aforementioned rescalings it is straightforward to evaluate it 
numerically in the $\sigma_* \to 0$ limit.
In Fig.~\ref{fig:NormalizedPhi}, we plot our result for $\Phi(x,x_\perp)$ in this limit.

\item As $x \to x_{\rm therm}$, when the string endpoint falls towards the horizon and we exit the SSR, the approximations used to derive 
(\ref{eq:holoflux2}) and obtain Fig.~\ref{fig:NormalizedPhi} all
break down.  For example, when $x_{\rm therm} - x \sim 1/T$, the near-endpoint geodesics curve downwards rapidly and are not well
approximated by the tangents (\ref{eq:nearboundarygeos}).  
Furthermore, as the string endpoint falls closer and closer to the horizon, the gravitational bulk to boundary problem cannot
be solved using zero temperature Green functions.    For any fixed small value of $\sigma_*$, meaning for any fixed
large value of $x_{\rm therm} T$, (\ref{eq:holoflux2}) and Fig.~\ref{fig:NormalizedPhi} are only good approximations as long
as $x_{\rm therm} - x \gtrsim 1/T$.  In the $\sigma_*\to 0$ limit, $x_{\rm therm} T\to \infty$ and the regime where they break down shrinks as a fraction
of $x_{\rm therm}$, and Fig.~\ref{fig:NormalizedPhi} is obtained in its entirety.

\item
In describing how the shape of the jet evolves as it propagates, we have focused entirely on
its expansion in the $x_\perp$ directions, transverse to its direction of motion $x$, as this
transverse expansion defines the opening angle of the jet.  That is, we have focused
on the expansion of the vertical dimension of the red oval in the cartoons in Figs.~\ref{fig:jetcartoon}
and \ref{fig:jetcartoon0}.  In our calculation, where we have worked to leading order in the $\sigma_*\to 0$ limit,
the red oval has zero longitudinal thickness. This can be seen by substituting the small $\sigma_*$ limits
for $\Delta t$ and $\Delta x$ given in (\ref{eq:DeltatDeltax}) into the expressions (\ref{eq:particlestress2})  and (\ref{eq:DeltaStress}) for
the near-jet stress-energy.   The delta function in (\ref{eq:particlestress2}) becomes $\delta(t - \Delta t - | \bm \Delta x|) = \delta(t - |\bm x|)$
which means that the longitudinal profile of the energy density depicted in the cartoons in Figs.~\ref{fig:jetcartoon}
and \ref{fig:jetcartoon0} is a delta function.  Of course,  (\ref{eq:DeltatDeltax}) is only valid in the SSR and 
to leading order in the  $\sigma_*\to 0$ limit.  We leave the calculation of the higher order corrections which
smear out the delta function to future work.

\end{enumerate}

\begin{figure}
\begin{centering}
\includegraphics[width=14cm]{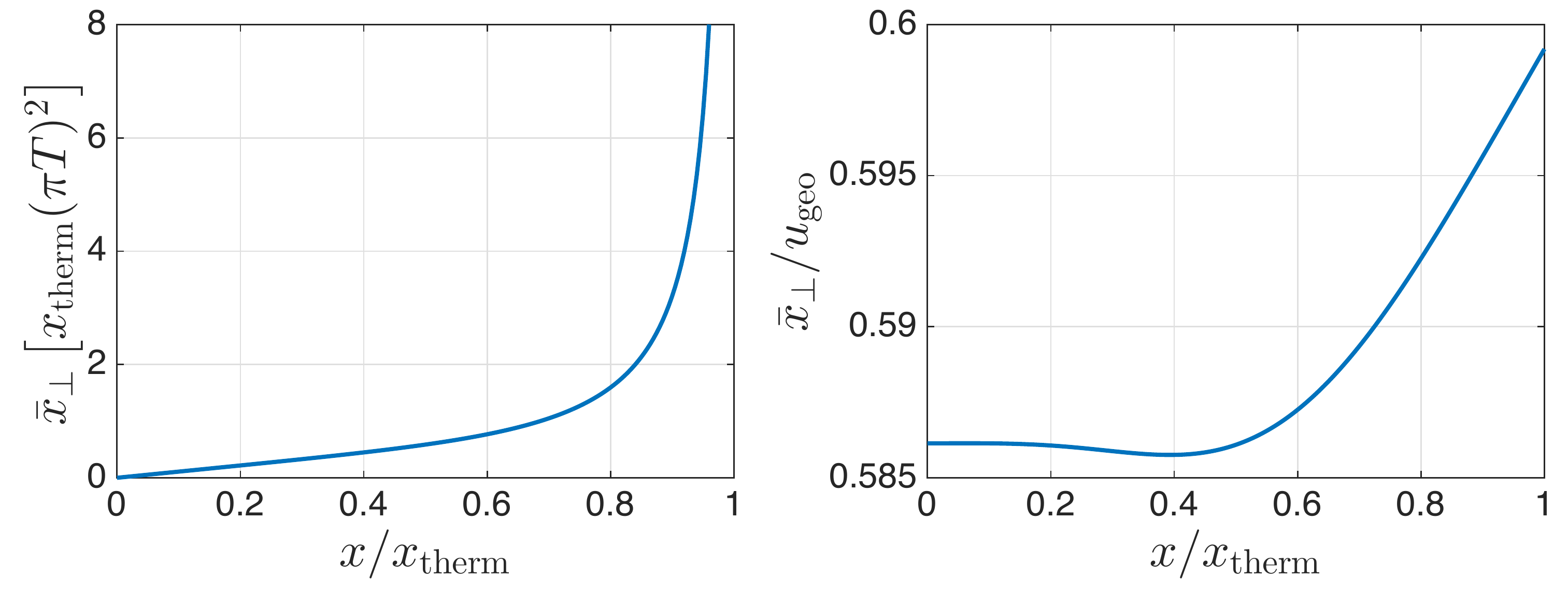}
\par\end{centering} 
\protect\caption{
\label{fig:TransverseEnvelope} 
Left: the normalized transverse envelope function $\bar x_\perp(x)$.
At small $x$ we have $\bar x_\perp \propto x$.  However, by $x/x_{\rm therm} \sim 0.5$
the rate of growth of $\bar x_\perp$ has begun to increase.  As $x$ increases further,
$\bar x_{\perp} \sim \frac{1}{T^2 x_{\rm therm}}\frac{1}{1 - x/x_{\rm therm}}$
until $x_{\rm therm} - x \sim 1/T$ at which point  $\bar x_\perp \sim 1/T$ and the jet thermalizes.
Right: the ratio of the transverse envelope function $x_\perp(x)$ to the radial coordinate of the string endpoint.
We see that $\bar x_\perp(x) \approx 0.59 \, u_{\rm geo}(\sigma_*,x)$ to within 1.5\% accuracy.   
This provides quantitative confirmation of a basic qualitative feature of the intuition underlying 
this application of the gauge/gravity correspondence,
namely that the transverse size of the jet in the boundary gauge theory is encoded in the dual gravitational
description  by how far the endpoint of the
string has fallen below the AdS boundary, down toward the black hole horizon.  The conclusion
that the opening angle of the jet is encoded by the downward angle of the black curve
in Fig.~\ref{fig:NullString} follows directly from this result.
}
\end{figure}

With $\Phi(x,x_\perp)$ in hand, we can now follow the prescription set out in Section~\ref{sec:elossdef}
to define the opening angle of the jet and follow its evolution as the jet propagates through  the strongly coupled plasma.
In the left panel of Fig.~\ref{fig:TransverseEnvelope}, we plot the transverse envelope function $\bar x_{\perp}(x)$, defined in Eq.~(\ref{eq:envelope})
as the half width at half maximum of $\Phi$.%
\footnote
  {
  With the explicit result (\ref{eq:holoflux2}) for $\Phi(x,x_\perp)$ available
  to us, we can now see why in Section~\ref{sec:elossdef} we chose to define the opening angle
  of the jet via $\bar x_{\perp}(x)$ rather than via a moment of $\Phi$, say 
  $\langle x_\perp(x) \rangle \equiv \frac{\int x_\perp^2 \Phi(x,x_\perp) d x_\perp}{\int x_\perp \Phi(x,x_\perp) d   x_\perp}$.
  Explicit evaluation shows that the leading order contribution to $\langle x_\perp \rangle$  in the $\sigma_*\to 0$ limit, i.e. in the
  small opening angle limit, is divergent.  The divergence reflects the fact that
  $\Phi(x,x_\perp)$ falls off so slowly at large $x_\perp$ that $\langle x_\perp(x) \rangle$ is parametrically larger than
  $\bar x_\perp$. In fact, as $\sigma_*\to 0$ and the jet opening angle defined from the half width at
  half maximum $\bar x_\perp$ scales to zero proportionally, the moment $\langle x_\perp(x) \rangle$
  instead stays nonzero.
  }  
As in Fig.~\ref{fig:NormalizedPhi}, in making the plot we have normalized  $\bar x_\perp$ by a factor of $x_{\rm therm}$ so
that we can plot a quantity that is fixed in the $\sigma_* \to 0$ limit.
The envelope function $\bar x_{\perp}(x)$ plotted in Fig.~\ref{fig:TransverseEnvelope} is of course a
constant-color contour of Fig.~\ref{fig:NormalizedPhi}.
At first $\bar x_\perp$ grows linearly in $x$ with rate 
\begin{align}
\label{eq:smallxxperp}
&\bar x_\perp = \beta \sigma_* x,&
&\beta \approx 0.5861519.
\end{align}
However, by $x/x_{\rm therm} \sim 0.5$ the rate of growth of $\bar x_\perp$ has begun to increase.  
Since $\hat u(\sigma_*,x)$ diverges as $x \to x_{\rm therm}$ \`a la Eqs.~(\ref{eq:xhatsigmah}) and (\ref{eq:uhatdivergence}), 
it follows that as $x \to x_{\rm therm}$  as long as the jet is within the SSR we must have the divergence 
\begin{equation}
\label{eq:widthdivergence}
\bar x_{\perp} \sim \frac{1}{T^2 x_{\rm therm}}\frac{1}{1 - x/x_{\rm therm}}.
\end{equation}
However, when $x_{\rm therm} - x \sim 1/T$ and the jet leaves the SSR,
we see from (\ref{eq:widthdivergence}) that 
the transverse size of the jet has grown to $\bar x_\perp \sim \frac{1}{T}$.

In the right panel of Fig.~\ref{fig:TransverseEnvelope} we plot $\bar x_\perp(x)/u_{\rm geo}(\sigma_*,x)$.
As is evident from the figure, the transverse envelope 
of the jet can be approximated by\footnote{
Note that if we had chosen to define $\bar x_{\perp}(x)$ as the half width at 10\% of maximum of $\Phi$ instead
of the half width at half maximum, which is to say if we had chosen a constant-color contour of Fig.~\ref{fig:NormalizedPhi}
in the deep-blue rather than in the mid-green, we would have obtained $\beta\approx 1.31009$ in (\ref{eq:smallxxperp})
and would have concluded here that $\bar x_\perp(x) \approx 1.34 \, u_{\rm geo}(\sigma_*,x)$ to within 2.5\% accuracy.
} 
\begin{equation}
\label{eq:envelopeapprox}
\bar x_\perp(x) \approx 0.59 \, u_{\rm geo}(\sigma_*,x)
\end{equation}
to within 1.5\% accuracy.
In other words, up to an ${\cal O}(1)$ normalization factor whose value depends
on an arbitrary choice of the definition of the width of the jet,
the radial coordinate of the string endpoint gives the transverse width of the jet. 
Indeed, using the small $\sigma$ limit of (\ref{eq:hatteddefs}) we see that the integrand in (\ref{eq:holoflux2})
is
\begin{equation}
\int_{\sigma_*}^{\sigma_h} d\sigma \frac{\hat u^4}{u_h^2 \left [  \sigma \hat u^2 +   x_{\perp}^2/u_h^2 \right ]^3} \frac{1}{ \sqrt{\sigma - \sigma_*}} = 
\int_{\sigma_*}^{\sigma_h} d\sigma \frac{ u_{\rm geo}^4}{ \left [  u_{\rm geo}^2 +   x_{\perp}^2 \right ]^3} \frac{1}{ \sigma^2 \sqrt{\sigma - \sigma_*}}.
\end{equation}
The width of the integrand is just $\sim u_{\rm geo}(\sigma,x)$ weighted by the world sheet energy (\ref{eq:Pi00UV}),
which diverges at the string endpoint.  Therefore, it is natural that $\bar x_\perp(x)$ is approximately proportional to 
$u_{\rm geo}(\sigma_*,x)$.

\begin{figure}
\begin{centering}
\includegraphics[width=14cm]{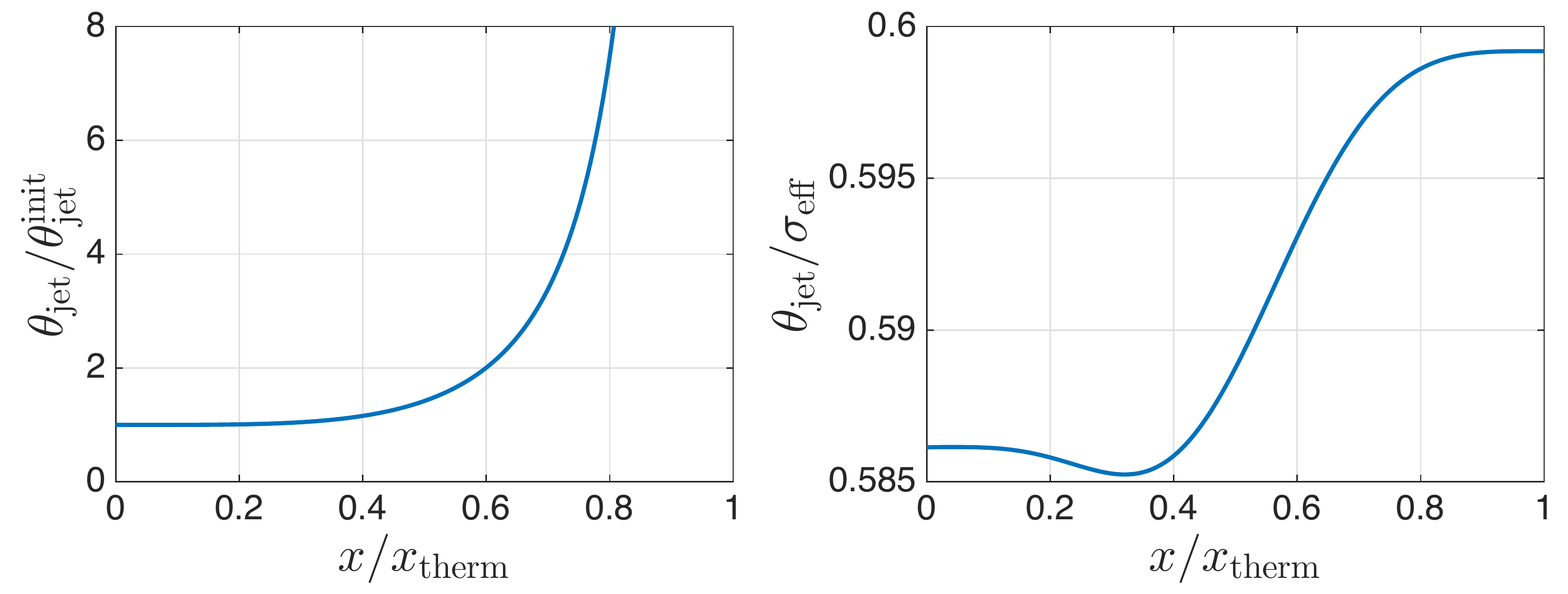} 
\par\end{centering} 
\protect\caption{
\label{fig:JetAngle} 
Left: the  jet opening angle $\theta_{\rm jet}$, in units of its initial value $\theta_{\rm jet}^{\rm init}$, 
as a function of $x/x_{\rm therm}$.  The opening angle is 
almost constant until $x/x_{\rm therm} \sim 0.5$.
As $x$ increases further, $\theta_{\rm jet}$ increases like 
$\theta_{\rm jet} \sim \frac{\theta_{\rm jet}^{\rm init}}{\left [1 - x/x_{\rm therm} \right ]^2},$
until $\theta_{\rm jet} = {\cal O}(1)$ when $x_{\rm therm} - x \sim 1/T$
and the jet thermalizes.  The figure is drawn in the limit $\sigma_*\to 0$, meaning
in the limit in which $\theta_{\rm init}\to 0$ and $x_{\rm therm}T\to\infty$; in this limit,
$\theta_{\rm jet}$  becomes of order 1 and the jet thermalizes when $x/x_{\rm therm}$ is arbitrarily close to 1.
Right: The ratio of the jet opening angle $\theta_{\rm jet}(x)$ to the angle $\sigma_{\rm eff}(\sigma_*,x)$ that the string endpoint trajectory
makes with the boundary.
We have  $\theta_{\rm jet} \approx 0.59\, \sigma_{\rm eff}$ for all $x/x_{\rm therm}$ to within 1.5\% accuracy.  
Therefore, the local jet opening angle 
is determined by the local angle the string endpoint trajectory makes with the AdS boundary, which is to say by the downward
angle of the black curve in Fig.~\ref{fig:nullstring}.  As gravity in the bulk makes the endpoint curve
downward towards the horizon, the opening angle of the jet on the boundary increases.
}
\end{figure}

In the small angle limit, the opening angle of the jet that we defined in  Eq.~(\ref{eq:thetadef0}) is given by 
\begin{equation}
\theta_{\rm jet} = \frac{d \bar x_{\perp}}{dx}.
\label{eq:smallthetajetfromxbarperp}
\end{equation}
In the left panel of Fig.~\ref{fig:JetAngle} we plot $\theta_{\rm jet}/\theta_{\rm jet}^{\rm init}$ as a function of $x/x_{\rm therm}$.
From (\ref{eq:smallxxperp}) it follows that at small $x$ the opening angle is almost constant, hardly changing from its initial value 
\begin{equation}
\label{eq:thetajetinitresult}
\theta_{\rm jet}^{\rm init} = \beta \sigma_*.
\end{equation}
Upon solving for $\sigma_*$ and plugging this result into the expression (\ref{eq:xthermsigma}) for the thermalization distance,
we obtain Eq.~(\ref{eq:xtherm2}) in the Introduction, namely
\begin{equation}
x_{\rm therm} = \frac{1}{ T} \sqrt{\frac{\kappa}{\theta^{\rm init}_{\rm jet}}},
\label{eq:xtherm2repeated}
\end{equation}
with 
\begin{equation}
\kappa = \frac{ \Gamma \left(\frac{1}{4}\right)^4}{16 \pi ^3}\,\beta \approx 0.204157.
\end{equation}
We now see that the total distance that a jet can travel through plasma before thermalizing 
is entirely determined by the temperature and the initial opening angle alone.  The initial
energy of the jet need not be known, although as we shall see in Section~\ref{sec:energyloss} it
must be greater than some minimal value that depends on the initial opening angle. 

Mirroring the behavior of $\bar x_\perp$, after initially increasing only very slowly, by 
$x/x_{\rm therm} \sim 0.5$ the rate of growth of $\theta_{\rm jet}$ has begun to increase. 
As $x$ increases further, as long as the jet is in the SSR we can see from  (\ref{eq:smallthetajetfromxbarperp}),
 (\ref{eq:widthdivergence}) and (\ref{eq:xtherm2repeated}) that $\theta_{\rm jet}$ increases like 
\begin{equation}
\label{eq:thetadivergence2}
\theta_{\rm jet} \sim \frac{\theta_{\rm jet}^{\rm init}}{\left[ 1 - \frac{{\textstyle x}}{\textstyle{x_{\rm therm}}}\right]^2}.
\end{equation}
When $x_{\rm therm} - x \sim 1/T$, the jet exits the SSR, our analysis breaks down, and the jet is no longer crisply defined.
This occurs when $\theta_{\rm jet}$ has increased to the point that
 $\theta_{\rm jet} = {\cal O}(1)$. 
We therefore conclude that the jet thermalizes when its opening angle is of order 1 and its transverse size
is of order $1/T$.

In the right panel of Fig.~\ref{fig:JetAngle} we plot the jet angle normalized by $\sigma_{\rm eff}(\sigma_*,x)$.
Again, we see that 
\begin{equation}
\theta_{\rm jet}(x) \approx 0.59 \, \sigma_{\rm eff}(\sigma,x),
\label{eq:thetaProptoSigmaeff}
\end{equation}
to within 1.5\% accuracy.
This simply follows from (\ref{eq:envelopeapprox}) and the small angle limit of the 
definition of $\sigma_{\rm eff}$ in (\ref{eq:sigmaeff}).  Via Eqs.~(\ref{eq:sigmaeff2}), (\ref{eq:hatuexpansion}), and (\ref{eq:sigmaHresult}),
we therefore obtain 
\begin{equation}
\frac{\theta_{\rm jet}(x)}{\theta_{\rm jet}^{\rm init}} \approx \sqrt{1 +\left[ F^{-1}\left( \frac{x}{x_{\rm therm}}\right)\right]^4}\,,
\end{equation}
stated in the Introduction in Eq.~(\ref{eq:mainopeningangleresult}).  
The expansion (\ref{eq:jetangleresultSmallx}) is then obtained 
by expanding this result in powers of $x/x_{\rm therm}$.  Equivalently, we can start by expanding the RHS of  (\ref{eq:geosol}) in powers of $\hat u$,
since $\hat u$ is small when $\hat x \sim x/x_{\rm therm}$ is small, obtaining
\begin{equation}
\hat x =\frac{\Gamma\left(\frac{1}{4}\right)^2}{4\sqrt{\pi}}  \, \hat u \left [ 1 
- \frac{1}{10}\hat u^4
+ \frac{1}{24}\hat u^8
- \frac{5}{208}\hat u^{12}
+\ldots \right ]\ ,
\end{equation}
then invert this series obtaining the small $\hat x$ expansion of $\hat u$, and then
substitute the resulting series into (\ref{eq:sigmaeff2}) and expand again, obtaining the small $\hat x$ expansion of $\sigma_{\rm eff}$ which,
using (\ref{eq:hatteddefs}), takes the form
\begin{equation}
\frac{\sigma_{\rm eff}(\sigma_*,x)}{\sigma_*} =   \left [1
+\frac{1}{2} \left(\frac{x\sqrt{\sigma_*}}{u_h}\right)^4
+\frac{3}{40} \left(\frac{x\sqrt{\sigma_*}}{u_h}\right)^8
+\frac{11}{1200} \left(\frac{x\sqrt{\sigma_*}}{u_h}\right)^{12}
+\ldots  \right ].
\end{equation}
Using (\ref{eq:thetaProptoSigmaeff}) and (\ref{eq:xthermsigma}) and defining $a$ as in (\ref{eq:adef}), we obtain (\ref{eq:jetangleresultSmallx}), as stated in the Introduction.

\begin{figure}
\begin{centering}
\includegraphics[width=15cm]{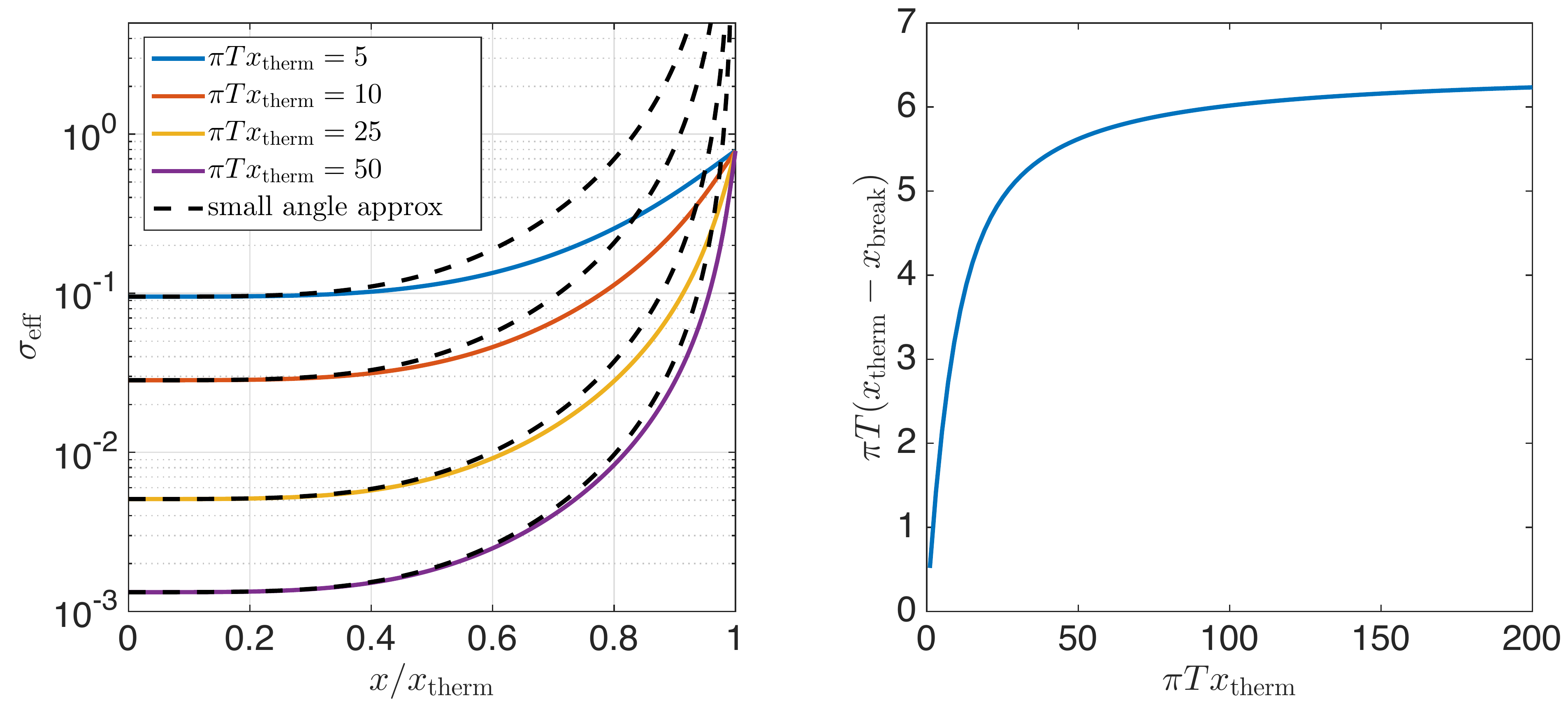} 
\par\end{centering} 
\protect\caption{
\label{fig:FiniteAngle} 
Left: the exact result for $\sigma_{\rm eff}(\sigma_*,x)$, given in (\ref{eq:sigmaeff1}), and the small
angle approximation for $\sigma_{\rm eff}(\sigma_*,x)$, given in (\ref{eq:sigmaeff2}), for 
$\sigma_*=0.0952$, 0.0284, 0.00509 and 0.00132
which yield the values of $x_{\rm therm}$ given in the legend.
The small-angle approximation works well until 
$x_{\rm therm}-x= {\cal O}(1/T)$.  
At larger $x$, beyond the SSR, all the colored curves converge to the common value $\sigma_{\rm eff}(\sigma_*,x_{\rm therm})\approx \pi/4$,  
corresponding to the fact discussed above that all the blue geodesics in the SSR in Fig.~\ref{fig:NullString} strike
the horizon at the same angle.  Although $\sigma_{\rm eff}$ remains well defined beyond the SSR, there is no well-defined jet
opening angle in this regime.
We can define $x_{\rm break}$ as the $x$ at which 
the small angle approximation for $\sigma_{\rm eff}(\sigma_*,x)$ is 25\% greater than
the exact result. Right: we see that $x_{\rm therm}-x_{\rm break}\sim 2/T$ in the small $\sigma_*$,
large $x_{\rm therm}T$, limit and is less than that at larger $\sigma_*$.
}
\end{figure}

We close this Section by comparing the exact result for $\sigma_{\rm eff}(\sigma_*,x)$, given in (\ref{eq:sigmaeff1}),
to the small-angle approximation of $\sigma_{\rm eff}(\sigma_*,x)$ given (\ref{eq:sigmaeff2}).  This comparison is useful
because it gives one an estimate of the domain of utility of our approximations for finite $x_{\rm therm}$.  How big is the SSR for 
some given $x_{\rm therm}$?  In Fig.~\ref{fig:FiniteAngle} we show the comparison between Eqs.~(\ref{eq:sigmaeff1}) and (\ref{eq:sigmaeff2}) at various values of 
$x_{\rm therm} T$.   We see from the figure that the small angle approximation works very well 
where $\sigma_{\rm eff}$ is almost constant and for the portion of its upward rise up until 
$x_{\rm therm}-x \sim 2/T$.  The range of $x$ where the colored curve peels away from its dashed
approximation therefore corresponds to the range of $x$ where the string leaves the SSR and our analysis of the energy loss rate 
and opening angle evolution breaks down.  Simply put, beyond this point, the instantaneous energy loss rate and opening angle are not crisply defined 
as the jet blurs as it thermalizes.

\subsection{Rate of energy loss}
\label{sec:energyloss}

We now perform the holographic computation of the rate at which the jet loses energy
to hydrodynamic modes of the plasma, 
which is to
say the rate at which energy flows through the yellow sphere in Fig.~\ref{fig:jetcartoon}.
As we discussed in Section~\ref{sec:elossdef}, in order to compute the rate of energy loss $dE_{\rm jet}/dt$
of the jet in the boundary gauge theory 
, we need the transverse traceless
mode of the stress tensor of the boundary gauge theory.  
From the relationship between the ${\cal N}=4$ SYM stress and the metric perturbation $H_{MN}$, Eq.~(\ref{eq:holorenorm0}),
it is clear that the transverse traceless mode of the SYM stress is encoded in the 
transverse traceless mode of $H_{MN}$, which in turn is sourced by the string stress tensor $J^{MN}$ that
we obtained in Eq.~(\ref{eq:stringstress}).  We now develop these relationships explicitly.

Let
\begin{equation}
\widetilde H_{MN}(\omega,\bm q) \equiv \int dt \, d^3 x \, H_{MN}(t,\bm x,u)  \> e^{i \omega t} e^{-i \bm q \cdot \bm x}
\end{equation}
be the Fourier transform of $H_{MN}(t,\bm x,u)$  and define the transverse traceless gravitational mode 
\begin{equation}
\label{eq:Zdef}
Z_{ab} \equiv \left [\epsilon_a^i \epsilon_b^j - {\textstyle \frac{1}{2}} \delta_{ab}  \epsilon_c^i \epsilon_c^j \right]   
\widetilde H_{i j}(\omega,\bm q,u).
\end{equation} 
It is then straightforward to show from the linearized Einstein field equations (\ref{eq:linEinstein}) that $Z_{ab}$ obeys the ordinary differential equation \cite{Chesler:2007sv}
\begin{equation}
\label{eqm2}
 Z''_{ab}+A \,  Z'_{ab} + B \,  Z_{ab} =  S_{ab} \,,
\end{equation}
where $' \equiv \partial_u$ and 
\begin{align}
\label{eq:odecoeffs}
& A \equiv \frac{u f' - 3f}{u f}\,, &
& B \equiv -\frac{q^2 f - \omega^2}{f^2} \,, &
& S_{ab} \equiv - \frac{16 \pi G_{\rm Newton}}{f}  
 \left [\epsilon_a^i \epsilon_b^j - {\textstyle \frac{1}{2}} \delta_{ab}  \epsilon_c^i \epsilon_c^j \right]  \widetilde J_{ij}, &
\end{align}
with $\widetilde J_{MN}$ the Fourier transform of the string stress tensor $J_{MN}$ and $q = |\bm q|$.
With the boundary condition  $\lim_{u \to 0} H_{MN} = 0$,
the ODE (\ref{eqm2}) implies that $Z_{ab}$ has  the near-boundary asymptotic behavior
\begin{equation}
\label{eq:Zasym}
Z_{ab}(\omega,\bm q,u) \sim u^4 Z^{(4)}_{ab}(\omega,\bm q) + O(u^5), 
\end{equation}
with 
\begin{equation}
Z^{(4)}_{ab}(\omega,\bm q) = \left [\epsilon_a^i \epsilon_b^j - {\textstyle \frac{1}{2}} \delta_{ab}  \epsilon_c^i \epsilon_c^j \right] \widetilde H_{ij}^{(4)}(\omega, \bm q).
\end{equation}
From the definition of the transverse traceless mode of the SYM stress in Eq.~(\ref{eq:transversetracelessdef}), and the 
relationship between $H_{\mu \nu}^{(4)}$ and $\langle \Delta T^{\mu \nu} \rangle$ in Eq.~(\ref{eq:holorenorm0}), 
we therefore conclude that the transverse traceless mode of the
boundary gauge theory stress tensor, defined in (\ref{eq:transversetracelessdef}), is given by
\begin{equation}
\label{eq:ttmodefromholo}
\mathcal T_{ab}(\omega,\bm q) = \frac{ L^3}{4 \pi G_{\rm Newton}} Z^{(4)}_{ab}(\omega,\bm q).
\end{equation}
Therefore, computing the long wavelength limit of 
$\mathcal T_{ab}$ is tantamount to solving (\ref{eqm2}) in the long wavelength limit.

To solve (\ref{eqm2}) we construct a Green's function $\mathcal G(\omega,\bm q,u,u')$ out of homogeneous 
solutions
$g_>$ and $g_{<}$,
\begin{equation}
\mathcal G(\omega,\bm q,u,u') = \frac{g_<(\omega,\bm q,u_<) g_>(\omega,\bm q, u_>)}{W(\omega,\bm q,u')},
\end{equation}
where $W$ is the  Wronskian of $g_<$ and $g_>$. The appropriate homogeneous solutions
are dictated by boundary conditions.  The differential operator in (\ref{eqm2}) has singular points
at $u = 0$ and $u = u_h$.  At $u=0$ the indicial exponents are $0$ and $4$ while at $u = u_h$ they 
are $\pm i\omega u_h/4$.  Vanishing of $Z_{ab}$ at the boundary requires $g_< \to 0$ as $u \to 0$, 
while the requirement that the black hole not radiate
requires $g_> \sim (u - u_h)^{-i\omega u_h/4}$ as $u \to u_h$.  We choose normalization so that
$g_{<} = u^4 + {\cal O}(u^5)$.  With this choice, the coefficient $Z_{ab}^{(4)}$ reads
\begin{equation}
\label{eq:z4def}
Z_{ab}^{(4)}(\omega,\bm q) = \int_0^{u_h} du \frac{g_>(\omega,\bm q,u)}{W(\omega,\bm q,u)} S_{ab}(\omega,\bm q,u).
\end{equation}
The integrand has two pieces, one coming from the homogeneous solutions and one coming from
the string stress tensor. We take them in turn.

In the long wavelength limit $\omega,q \to 0$, the homogeneous solutions to Eq.~(\ref{eqm2}) can be computed analytically and read $g_< = -u_h^4 \log f$
and $g_> = 1.$  We therefore secure the long wavelength asymptotic behavior  
\begin{equation}
\label{eq:longwavelengthgravsol}
\frac{g_>(\omega,\bm q,u)}{W(\omega,\bm q,u)} =  -\frac{f}{4 u^3} + { O}(q),
\end{equation}
that is the first ingredient needed in order to evaluate (\ref{eq:z4def}) and hence (\ref{eqm2}) at long wavelengths.

Next, we need $S_{ab}$. We choose polarization vectors
\begin{align}
\label{eq:polarziationvecs}
&\bm \epsilon_1 = \frac{q}{q_{\perp}} \hat { q} \times ( \hat { x} \times  \hat { q}),&
\bm \epsilon_2 = \frac{q}{q_{\perp}} \hat { x} \times \hat { q},&
\end{align}
with $\hat q = \bm q/q$ and $q_\perp = |\bm q - (\hat {x} \cdot \bm q) \hat { x} |$.  
Fourier transforming the string stress (\ref{eq:stringstress}),
we obtain
\begin{equation}
\label{eq:FTstringStress}
\widetilde J_{ij}(\omega,\bm q,u) = -   \frac{u^3}{L^3} \int d\sigma \frac{\hat x^i \hat x^j}{\xi(\sigma) \sqrt{\xi(\sigma)^2 - f(u)}} \,
\Pi^\tau_0(\sigma)\, e^{  i \omega t_{\rm geo}(\sigma,u) - i q_x x_{\rm geo}(\sigma,u)},
\end{equation} 
where $t_{\rm geo}(\sigma,u)$ is given by the solution to $u_{\rm geo}(t = t_{\rm geo},\sigma) = u$
and $x_{\rm geo}(\sigma,u)$ is given by Eq.~(\ref{eq:geosol}).
Using Eqs.~(\ref{eq:FTstringStress}), (\ref{eq:polarziationvecs}) and (\ref{eq:odecoeffs}),
the source $S_{ab}$ reads
\begin{subequations}
\label{eq:explicitSourceComponents}
\begin{align}
S_{11}(\omega,\bm q,u) &=  \frac{ 8 \pi G_{\rm Newton} u^3}{L^3 f(u)} \left (\frac{q_{\perp}}{q} \right)^2 \int d\sigma \frac{ 
\Pi_0^\tau(\sigma) }{\xi(\sigma)  \sqrt{\xi(\sigma)^2 - f(u)}} e^{i \omega t_{\rm geo}(\sigma,u) - i  q_{x} x_{\rm geo}(\sigma,u)}, \\
 S_{2a}(\omega,\bm q,u) &= S_{a2}(\omega,\bm q,u) = 0.
\end{align}
\end{subequations}

We now have all the pieces needed
to evaluate (\ref{eq:z4def}).
Using the long wavelength solution (\ref{eq:longwavelengthgravsol}) and the source (\ref{eq:explicitSourceComponents}), we conclude from 
(\ref{eq:z4def}) that the long wavelength limit of the coefficient $Z^{(4)}_{ab}$ is given by 
\begin{subequations}
\begin{align}
\label{eq:nonzeroZ4}
Z_{11}^{(4)}(\omega,\bm q) &= -\frac{2 \pi G_{\rm Newton}}{ L^3}\left (\frac{q_{\perp}}{q} \right)^2 \int d\sigma \frac{\Pi_0^\tau(\sigma)}{\xi(\sigma)} \int_0^{u_h} du\frac{ 
1}{\sqrt{\xi(\sigma)^2 - f(u)}} e^{i \omega t_{\rm geo}(\sigma,u) - i  q_{x} x_{\rm geo}(\sigma,u)}, \\
 Z^{(4)}_{2a}(\omega,\bm q) &=Z^{(4)}_{a2}(\omega,\bm q) =0.
 \end{align}
 \end{subequations}
 
We wish to evaluate the radial integral in (\ref{eq:nonzeroZ4}) in the limit $x_{\rm therm} \ggg 1/T$ with
$\omega x_{\rm therm}$ and $q_x x_{\rm therm}$ fixed.  The integral is dominated by $u \sim \sqrt{\sigma} u_h$.
In the region $u \sim \sqrt{\sigma} u_h$ the geodesic equations (\ref{eq:geo1}) imply $t_{\rm geo} \approx x_{\rm geo}$.
Furthermore, we see from the geodesic equations (\ref{eq:geo1})  that $\frac{dx_{\rm geo}}{du} = \frac{1}{\sqrt{\xi^2 - f}}$.
This means  the radial integrand in (\ref{eq:nonzeroZ4}) reads
\begin{equation}
\frac{ 
1}{\sqrt{\xi(\sigma)^2 - f(u)}} e^{i \omega t_{\rm geo}(\sigma,u)   - i q_{x} x_{\rm geo}(\sigma,u)} \approx \frac{1}{i (\omega - q_x)} \frac{\partial}{\partial u} e^{i (\omega -  q_{x}) x_{\rm geo}(\sigma,u)} \, .
\end{equation}
Therefore, in the IR the radial integral in (\ref{eq:nonzeroZ4}) evaluates to 
\begin{equation}
\label{eq:longwavelengthradialintegral}
\int_0^{u_h} du\frac{ 
1}{\sqrt{\xi(\sigma)^2 - f(u)}} e^{i \omega t_{\rm geo}(\sigma,u) - i  q_{x} x_{\rm geo}(\sigma,u)}
= \frac{1}{i (\omega - q_x)} \left [ e^{i (\omega -  q_{x}) x_{\rm stop}(\sigma)} - 1 \right ],
\end{equation}
where we have used $x_{\rm geo}(\sigma,u = u_h) = x_{\rm stop}(\sigma)$.

Using (\ref{eq:longwavelengthradialintegral}), (\ref{eq:nonzeroZ4}) and (\ref{eq:ttmodefromholo}), we therefore 
conclude that the long wavelength limit of the transverse traceless mode of the gauge theory stress tensor reads
\begin{align}
\label{eq:holoresult}
&\mathcal T_{11}^{(4)}(\omega,\bm q)  =   \frac{1}{2} \left (\frac{q_\perp}{q} \right )^2 \frac{1}{i (\omega - q_x)}  \int d\sigma \, \Pi^\tau_0(\sigma) [1-e^{i (\omega-q_x) x_{\rm stop}(\sigma)}],
&
\mathcal T_{2 a}  =\mathcal T_{a2}= 0.&
\end{align}
We have achieved our goal; what remains is interpreting (\ref{eq:holoresult}) and understanding its consequences.

Eq.~(\ref{eq:holoresult}),
should be compared to the expected form in Eq.~(\ref{eq:elossfromTTmode}).  Substituting  both $\bm V \approx \hat x$ and the polarization vectors (\ref{eq:polarziationvecs}) into 
(\ref{eq:elossfromTTmode}), and comparing to (\ref{eq:holoresult}), we see that the jet energy $\widetilde E_{\rm jet}$ must be given by 
\begin{equation}
\widetilde E_{\rm jet}(\omega) = \frac{1}{i \omega} \int d\sigma \, \Pi^\tau_0(\sigma) [1-e^{i \omega x_{\rm stop}(\sigma)}].
\end{equation}
Fourier transforming back to real space, the leading order derivative expansion of the jet energy reads 
\begin{equation}
\label{eq:jetenergyresult}
E_{\rm jet}(t) = - \int d\sigma \, \Pi_0^\tau(\sigma) \left[\theta(t) - \theta(t - x_{\rm stop}(\sigma)\right].
\end{equation}
The first term in (\ref{eq:jetenergyresult})  just corresponds to the energy added when the jet is created
while the second encodes the energy loss of the jet while it propagates through the plasma.
Using the fact that the jet moves at nearly the speed of light,  the energy lost per unit distance traveled reads
\begin{equation}
\label{eq:jetenergylossrateresult}
\frac{dE_{\rm jet}}{dx} = - \Pi^\tau_0(\sigma_h) \frac{d\sigma_h}{dx},
\end{equation}
where $\sigma_h$ is given in (\ref{eq:sigmaHresult}).
In the dual gravitational picture, the energy loss rate (\ref{eq:jetenergylossrateresult}) is nothing more than the rate at which
the string's energy flows into the horizon.  
This was introduced as a definition in Ref.~\cite{Chesler:2014jva}.
We now see that this is indeed the natural definition, as it is equivalent to defining the energy loss 
rate in the field theory as the rate at which energy flows into hydrodynamic modes, which is to
say the rate at which energy flows through the yellow sphere in Fig.~\ref{fig:jetcartoon}.

Using Eqs.~(\ref{eq:Pi00UV}), (\ref{eq:sigmaHresult}) and (\ref{eq:xthermsigma}), it is a 
straightforward exercise to show that (\ref{eq:jetenergylossrateresult})
yields 
\begin{equation}
\label{eq:energylossresult0repeated}
\frac{1}{E_{\rm init}} \frac{dE_{\rm jet}}{dx} = - \frac{4 x^2}{\pi x_{\rm therm}^2 \sqrt{x_{\rm therm}^2 - x^2}},
\end{equation}
our result for the rate of energy loss, given in the Introduction as Eq.~(\ref{eq:energylossresult0}), discussed
there extensively, and derived
previously in Ref.~\cite{Chesler:2014jva},
with the initial energy given by
\begin{equation}
\label{eq:initialenergy}
E_{\rm init} = \frac{\sqrt{\lambda}}{4\sqrt{2\epsilon\, \psi_1} } \frac{1}{\sigma_*^{3/2}}\,,
\end{equation}
as can also be seen by evaluating (\ref{eq:jetenergyresult}) at $t=0^+$ using (\ref{eq:Pi00UV}).

As in our calculation of the evolution of the opening angle of the jet
in Section~\ref{sec:openingangle}, these expressions are valid only in the SSR.
In the $x_{\rm therm}T\to\infty$ limit, the SSR extends
to  $x/x_{\rm therm}$ arbitrarily close to 1, meaning that 
(\ref{eq:energylossresult0repeated}) can be used until $E_{\rm jet}/E_{\rm init}$ is
arbitrarily close to 0.  
For jets with a large but finite $x_{\rm therm}T$, meaning
a small but nonzero initial opening angle, (\ref{eq:energylossresult0repeated})
breaks down where $x_{\rm therm}-x\sim 1/T$ and the jet leaves the SSR.
$E_{\rm init}$ is the energy of the jet when it enters the SSR. 
It is impossible to define
the jet energy from the distribution of energy along the string before
the string enters the SSR because at the moment of its creation the string
describes gluon fields around its creation event as well as the jet, and we need
to wait for the jet to separate from extraneous, transient, gluonic excitations which
are not comoving with the jet and which therefore fall 
into the horizon within a time of order $1/T$ before defining $E_{\rm jet}$.
In the $x_{\rm therm}T\to\infty$ limit, the jet enters the SSR
at a value of $x/x_{\rm therm}$ that is arbitrarily close to 0, and the 
result (\ref{eq:energylossresult0repeated}) can be applied for $0< x/x_{\rm therm} < 1$ and
describes the evolution of $E_{\rm jet}$ from $E_{\rm init}$ to 0.

Upon solving (\ref{eq:initialenergy}) for $\sigma_*$ and substituting into 
(\ref{eq:xthermsigma}) and (\ref{eq:thetajetinitresult}),  we obtain 
the relationship between the initial energy and both the thermalization distance and the initial opening angle:
\begin{align}
&x_{\rm therm} = \frac{1}{T} \left ( \frac{E_{\rm init}}{E_0} \right )^{1/3}, &
&\theta_{\rm jet}^{\rm init} = \kappa  \left ( \frac{E_{\rm init}}{E_0} \right )^{-2/3},
\end{align}
both results that we gave in the Introduction, namely Eqs.~(\ref{eq:thermalizationdistance0}) and (\ref{eq:openingangle}), now with
an explicit expression for the energy scale $E_0$:
\begin{equation}
E_0\equiv \frac{1}{\Gamma\left({\textstyle \frac{1}{4}}\right)^6}\left( \frac{128\, \pi^9 \lambda}{\epsilon\, \psi_1}\right)^{1/2}\,.
\label{CurlyC}
\end{equation}
The preparation of the initial state of the string enters $E_0$, and hence our 
results  (\ref{eq:openingangle}) and (\ref{eq:thermalizationdistance0}), only 
through the constant $\psi_1$ defined in (\ref{eq:psiexpansion}).   
We see from (\ref{eq:openingangle}) and (\ref{CurlyC}) that $E_{\rm init}$ is fully specified by $\theta^{\rm init}_{\rm jet}$ and
$\psi_1$.  Through (\ref{eq:xtherm2}) we see that $\theta^{\rm init}_{\rm jet}$ and $T$ specify $x_{\rm therm}$ meaning that 
(through (\ref{eq:energylossresult0}) 
or (\ref{eq:energylossresultSmallx}) and (\ref{eq:mainopeningangleresult}) or (\ref{eq:jetangleresultSmallx}))
$\theta^{\rm init}_{\rm jet}$, $\psi_1$ and $T$ fully specify the rate of energy loss
and the evolution of the jet opening angle as the jet propagates through the plasma.

What is the gravitational origin of the divergence in $\frac{dE_{\rm jet}}{dx}$ as $x \to x_{\rm therm}$
that is apparent in (\ref{eq:energylossresult0repeated})?
The opening string boundary conditions require that the string endpoint moves at the speed of light.
This in turn means that $\Pi^\tau_0(\sigma)$ must diverge at the endpoint, as 
Eq.~(\ref{eq:Pi00UV}) demonstrates, since the string has finite tension.  As $x \to x_{\rm therm}$, the flux of energy into the horizon 
must increase because more and more of the near-endpoint geodesics fall into the horizon.
Simply put, the dramatic increase in the energy loss rate as $x \to x_{\rm therm}$ encodes the fact 
that the string endpoint energy density is divergent and the last to fall into the horizon.

The relationship between the initial opening angle and the initial energy of the jet depends
on the details of how the state is prepared, details that are encoded in the value of $E_0$.
We can now see why $E_0$ must have a minimum value.  From (\ref{CurlyC}) we see that 
decreasing $E_0$ corresponds to increasing $\epsilon\, \psi_1$.
However, this cannot be done without bound, since the geometric optics expansion (\ref{eq:stringexpansion}) will eventually break down.  
Equating derivatives of the zeroth and first order terms in the geometric expansion, 
$\partial_\sigma x_{\rm geo} \sim \epsilon \partial_\sigma \delta x_{(1)}$,
via Eqs.~(\ref{eq:hatteddefs}), (\ref{eq:geosol}), (\ref{eq:Pi00}) and (\ref{eq:Pi00UV}) we estimate that the geometric optics expansion breaks down when 
$\epsilon \,\psi_1 \sim 1/T^2$, which means ${\rm min}(E_0) \sim \sqrt{\lambda} T$.
Equivalently, this means that the dimensionless constant $\mathcal C$ introduced in the Introduction via
$E_0= T\sqrt{\lambda}/{\mathcal C}^3$ has a maximum value which is of order 1.  Indeed, 
both analytical arguments and numerical solutions to the string equation of motion suggest that 
the maximum value of $\mathcal C$ ranges between 0.5 and 1 
\cite{Chesler:2008uy,Ficnar:2013wba}.
The fact that $E_0$ has a minimum possible value corresponds to the statement
that jets with a given $\theta_{\rm jet}^{\rm init}$ can have any initial energy $E_{\rm init}$ above some minimum possible value,
or to the statement that jets with a given $E_{\rm init}$ must have initial opening angles greater than some minimum possible value.
Knowing the initial energy of a jet requires knowing both its initial opening angle and the details of how it was prepared
that are encoded in $\psi_1$ and hence $E_0$.
Therefore, the relationship between the initial energy of the jet and its thermalization distance also depends on these details.  
But, the relationship between the initial
opening angle of the jet and its thermalization distance is independent of any details concerning how the state is prepared; it is fully
specified by (\ref{eq:xtherm2}).  This is one of the central lessons we learn from our calculation.

While we have obtained the energy loss rate from the rate that energy flows into IR modes, we 
note that we could have equally well obtained the energy loss rate from the rate that energy leaves UV modes.
Specifically, it must be that 
\begin{equation}
\frac{dE_{\rm jet}}{dt} = \frac{d}{dt} \int d^3 x \langle T^{00}_{\rm near-jet} \rangle,
\end{equation}
with $\langle T^{00}_{\rm near-jet} \rangle$ given in (\ref{eq:DeltaStress}).  Indeed, a straightforward 
calculation shows this to be the case.  

\section{Qualitative lessons for jets in heavy ion collisions}
\label{sec:concludingremarks}

We have provided a complete summary of all of our central results, in their own context, in the Introduction.
That is, we described the conclusions that we draw from
our results and their many consequences and implications for the behavior of the jets in ${\cal N}=4$ SYM theory
that we have analyzed.  We shall not repeat these  conclusions and observations here.  
Our purpose in this Section is to speculate as to how one may draw qualitative lessons for
jets in QCD, as produced and subsequently 	quenched in heavy ion conclusions, from our results.

As we already noted in Section~\ref{sec:Introduction}, there are two broad paths possible here. The more
conservative approach is to observe that perturbative QCD does a fine job of describing the high-momentum-transfer 
physics of jet production and jet fragmentation, and conclude that we should use insights gleaned from calculations
in strongly coupled ${\cal N}=4$ SYM theory only to  guide how we treat the low-momentum-transfer interactions
between the individual partons within a QCD jet shower and the strongly coupled quark-gluon plasma in which they find themselves.
The first steps down this path have been taken in Refs.~\cite{Casalderrey-Solana:2014bpa,Casalderrey-Solana:2015vaa}, with the construction of a hybrid model that
utilizes the form of the expressions (\ref{eq:energylossresult0}) and (\ref{eq:thermalizationdistance0}) that we have rederived
for the rate of energy loss $dE_{\rm jet}/dx$ of  ${\cal N}=4$ SYM jets
to describe the rate of energy loss of partons in a shower that is otherwise as described by perturbative
QCD.  Our work supports this approach via rederiving (\ref{eq:energylossresult0}) and (\ref{eq:thermalizationdistance0}) without
assuming a slab of plasma and with $dE_{\rm jet}/dx$ defined entirely within the gauge theory.  It is interesting to ask how
our results concerning the opening angle of the ${\cal N}=4$ SYM theory jets that we have analyzed can inform future extensions of
the hybrid model of Refs.~\cite{Casalderrey-Solana:2014bpa,Casalderrey-Solana:2015vaa}. 
Its simplest incarnation which has been used to this point relies upon (\ref{eq:thermalizationdistance0})
with
a single, in effect average, value of $E_0/(T\sqrt{\lambda})$ that has been  
obtained for jets in quark-gluon plasma via a fit to heavy ion collision data.  From (\ref{eq:openingangle}) we now
see that for a given $E_{\rm init}$ the constant $E_0$ is different for jets with different initial opening
angles.  So, possible extensions to the hybrid model could either introduce a probability distribution
for $E_0$ reflecting the distribution of initial opening angles or could directly relate $E_0$ to the initial  energy
and initial opening angle of each jet in a Monte Carlo sample
according to (\ref{eq:openingangle}), in effect introducing an initial-opening-angle-dependent $E_0$ rather than fitting a single
average value in units of $T\sqrt{\lambda}$.  After extending the model in this way one could then
use (\ref{eq:xtherm2}) to fix $x_{\rm therm}$ for each jet after which the energy lost by each
parton in that jet could be described by (\ref{eq:energylossresult0}) as in Refs.~\cite{Casalderrey-Solana:2014bpa,Casalderrey-Solana:2015vaa}.
In implementing this procedure, 
the value of the constant $\kappa$ in (\ref{eq:openingangle}) and (\ref{eq:xtherm2}) could be allowed to float, fitting
it to data in order to incorporate the differences between quark-gluon plasma and ${\cal N}=4$ SYM plasma.
Another possible extension
would be to somehow encode our result (\ref{eq:jetangleresultSmallx}) for how the opening angle of the ${\cal N}=4$ SYM jets
increases as they propagate in a probability distribution for the transverse kicks
that the partons in a perturbative QCD shower receive as they propagate through strongly coupled quark-gluon plasma, adding
the effects of such kicks to the hybrid model of Refs.~\cite{Casalderrey-Solana:2014bpa,Casalderrey-Solana:2015vaa}.

The more ambitious approach is to try to use insights gleaned from our results for ${\cal N}=4$ SYM jets to
obtain qualitative insights into the behavior of QCD jets in their entirety, without relying on a perturbative QCD
parton-by-parton description of the jets at all.  In the remainder of this Section, we look 
toward a variety of possible comparisons between the behavior of ${\cal N}=4$ SYM jets
and the behavior of jets in heavy ion collisions, as described perturbatively or as measured
in experiments:

\begin{itemize}

\item
It is interesting that, as we explain in full in Appendix~\ref{app:JetMass}, we find that the ratio of the initial jet mass to the initial
jet energy  $M_{\rm init}/E_{\rm init}$ is not a good proxy for the jet opening angle $\theta_{\rm jet}$,
because it 
is sensitive to the contribution of the   ``tails'' of the jets at angles
that are substantially greater than $\theta_{\rm jet}$.
This supports the use of measures of the jet opening angle that, like the half width at half maximum definition 
of $\theta_{\rm jet}$  that we have employed,
are defined from the jet shape, a quantity which has been measured in
heavy ion collisions~\cite{Chatrchyan:2013kwa}.   {Via suitable modelling in a Monte Carlo
study, it may also be possible to relate such measures to the ratios of the inclusive cross-sections
for the production of jets reconstructed from experimental data using different values of the radius parameter $R$ in the anti-$k_T$ reconstruction
algorithm
as in Refs.~\cite{Adam:2015doa,Jacobs:2015srw}.}

\item
Perhaps our most interesting result is the relationship (\ref{eq:xtherm2}) between
$x_{\rm therm}$ and $\theta_{\rm jet}^{\rm init}$. Together with (\ref{eq:energylossresult0}),
this tells us that jets with a smaller (larger) initial opening angle lose less (more) energy.
At a qualitative level, this is also certainly the case for jets in perturbative QCD, for the simple
reason that in perturbative QCD a jet with a larger initial opening angle is a jet that has
fragmented {into more partons and in particular into more resolved subjet structures,} each of which loses energy as it passes through the plasma,
meaning that a large-angle jet loses more energy in sum than a narrow jet containing
fewer partons does~\cite{CasalderreySolana:2012ef}.  
It is in fact very important to keep this feature in mind
in analyzing jet data.  If we compare two samples of jets in
heavy ion collisions that have, on average, lost different amounts of
energy --- for example a sample of leading jets (those that are the most
energetic jets in their event)  and a sample of jets that are the lower-energy jets
in a dijet pair --- then the leading jets will on average be those
which had a smaller initial opening angle.  The importance
of this effect has recently been emphasized~\cite{GuilhermeKorinna,Casalderrey-Solana:2015vaa}.
(The leading jets will also on average be those
which have travelled through a shorter length of plasma, but
Monte Carlo studies suggest that this may be the less important
effect~\cite{GuilhermeKorinna}.)

\item
Although narrow jets lose less energy than wide
jets both in ${\cal N}=4$ SYM theory and in perturbative QCD, there {can certainly be}
qualitative differences.  For example, it is striking that, as we discussed
in Section~\ref{sec:Introduction} via our results (\ref{eq:xtherm2}) and (\ref{eq:energylossresult0}),
the initial opening angle  of the ${\cal N}=4$ SYM jets controls their {\it fractional}
energy loss $\Delta E/E_{\rm init}$.  
{It would be interesting to investigate whether this qualitative regularity applies also to jets} in perturbative QCD, 
{but this is not expected since} 
in perturbative calculations of parton energy loss one obtains a $\Delta E$ for each {parton (more precisely, each resolved subjet)} that is 
independent of its initial energy or depends on it only logarithmically~\cite{Baier:1996kr,Baier:1996sk,Gyulassy:2000fs,Gyulassy:2000er,Mehtar-Tani:2013pia}.

\item
Our results make it clear that it would be exceedingly interesting to
tag jets in heavy ion collisions according to what their opening angle
would have been if they were produced in vacuum.  Unfortunately, we
know of no way to do this.  One can to some degree tag jets by what
their energy would have been if they were produced in vacuum by
looking at jets produced back-to-back with a photon or Z-boson whose
energy is measured.  But, we do not know of any way to know
what the opening angle of a particular jet seen in a heavy ion collision would have
been if it had not interacted with any quark-gluon plasma, i.e.~if it
had instead been produced in an elementary collision in vacuum.
This makes it particularly important to attempt a comparison between
the qualitative features of the relationship between
$\theta_{\rm jet}^{\rm init}$ and energy loss in (\ref{eq:xtherm2})  and 
(\ref{eq:energylossresult0}) to that same relationship in Monte Carlo
implementations of jet quenching in perturbative QCD, like for
example JEWEL~\cite{GuilhermeKorinna,Zapp:2013vla}.

\item
In any future comparison between (\ref{eq:xtherm2}) and (\ref{eq:energylossresult0})
on the one hand and information about the relationship between jet angle and jet
energy loss in QCD, either from Monte Carlo studies or from experiment, it will
be important to treat the purely numerical prefactor $\kappa$ 
in (\ref{eq:xtherm2}) as a free parameter.
If  the relationship (\ref{eq:xtherm2}) turns out to be a good description of the behavior of jets
in QCD, the value of the purely numerical prefactor must be different in QCD
than in ${\cal N}=4$ SYM theory because the  plasmas of the two theories have
different degrees of freedom.

\item
The difficulty in identifying what the opening angle of a particular
jet seen in a heavy ion collision would have been in vacuum makes
it hard to compare the most striking qualitative features of our results 
to experimental data in a direct fashion.  It should nevertheless be possible
to make such a comparison, albeit slightly more indirectly. One path
to a comparison with experimental
data is to (i) create an ensemble of ${\cal N}=4$ SYM jets
with varying $E_{\rm init}$ and $\theta_{\rm jet}^{\rm init}$,
perhaps choosing the distribution of $\theta_{\rm jet}^{\rm init}$
for jets with a given $E_{\rm init}$ following results from
perturbative QCD; (ii) then define $x_{\rm therm}$ for each of the jets
in the ensemble from $\theta_{\rm jet}^{\rm init}$ according
to (\ref{eq:xtherm2}), treating 
$\kappa$ 
as a 
free parameter to be varied; and (iii) then send this ensemble of jets through
a length of plasma that is small compared to all of their $x_{\rm therm}$'s.
With this setup, the decrease in the energy of each jet can be
obtained from (\ref{eq:energylossresultSmallx}) and the increase in the opening angle of each
jet can be obtained from (\ref{eq:jetangleresultSmallx}). 
Because the expansion (\ref{eq:jetangleresultSmallx}) starts at 
a higher power of $x/x_{\rm therm}$ than the expansion (\ref{eq:energylossresultSmallx}),
we expect that in a regime in which 
the fractional energy loss is small  the growth in the jet opening angle
will be even smaller.  Nevertheless, because the probability distribution
for $E_{\rm init}$ is steeply falling and that for $\theta_{\rm jet}^{\rm init}$ is
nontrivial, the effects of passing the ensemble of jets  
through the plasma on these distributions could be nontrivial.
It would be very interesting to
see how the final double-distribution, as a function of $E_{\rm jet}$ and $\theta_{\rm jet}$,
compares to the initial double-distribution, 
to ask under what circumstances the final distribution looks like a scaled version
of the initial distribution,
and to compare (suitable integrals of) the
final double-distribution to data.  

\item
Although perhaps less relevant to data, since what experimentalists {typically} see are jets
that emerge from a heavy ion collision as jets not jets that have thermalized,
it is also interesting to look at the qualitative behavior of ${\cal N}=4$ SYM jets
as they thermalize.  They lose a large fraction of their
energy over the last small fraction of their lifetime, reminiscent of
a Bragg peak~\cite{Chesler:2008uy,Chesler:2014jva}.  And, at the
end of their lifetime as
they thermalize their opening angle grows rapidly, until it is of order 1, 
and their transverse size also grows rapidly, until it is of order $1/T$.  
The question of whether analogous behavior is seen in 
jets in perturbative QCD could be investigated in Monte Carlo studies.
{Although challenging, it may also be possible to investigate such behavior via measuring jets 
as correlated energy flow over regions of large jet radius in events selected by
triggering on a photon or Z boson.}

\end{itemize}

At present it is too soon to tell how the many interesting qualitative features of our
results for the behavior of the energy and opening angle of ${\cal N}=4$ SYM jets
compare to experimental data or to calculations based upon perturbative QCD.
Such comparisons will ultimately determine how valuable it is to compare 
${\cal N}=4$ SYM jets to QCD jets, which is to say how many qualitative lessons
from the former apply to the latter.  Even if such comparisons remain as challenging
as they are at present or turn out unsuccessfully, though, the use of our results
in (future generalizations of) the hybrid strong/weak coupling model of 
Refs.~\cite{Casalderrey-Solana:2014bpa,Casalderrey-Solana:2015vaa}
will remain important, given how successfully the simplest one-parameter version
of this model has been able to describe so much jet data to date.

\vspace{0.5in}
\noindent{\bf Note Added:} 
\vspace{0.15in}

{
\noindent Since v1 of this paper was posted, in Ref.~\cite{Milhano:2015mng} Milhano and Zapp have completed the work 
that we have cited as a private communication~\cite{GuilhermeKorinna}.  They conclude from their 
analysis using the JEWEL event generator~\cite{Zapp:2013vla} that the dijet asymmetry
seen in LHC heavy ion collisions depends largely on the asymmetry between the
initial opening angles of the two jets and only to a subleading degree on the asymmetry
between the lengths of plasma which the two jets traverse.  As they note, their
results --- obtained in a calculation done entirely upon assuming weak coupling ---
are in qualitative agreement with our result --- obtained in a calculation done entirely upon assuming strong
coupling --- that the fractional jet energy loss
is controlled by the initial jet opening angle.
}

{Also since v1 of this paper was posted, in Ref.~\cite{Casalderrey-Solana:2015tas} Casalderrey-Solana
and Ficnar have presented beautiful results from a holographic calculation of three-jet events
in strongly coupled ${\cal N}=4$ SYM plasma.  They analyze classical string configurations
with non-trivial transverse dynamics in the initial state.  They find that their initial wave on
the string develops into a kink-like structure which can correspond in the gauge theory
to a three-jet event, with the string endpoints corresponding to quark jets and the kink corresponding 
to a gluon jet. Their study is the first analysis of proxies for three-jet events in a holographic context.
One of their central results can be described in terms of
results that we have obtained by observing that
their quark  and gluon jets propagate through the plasma as two separate jets if and only
if the angular separation between them is greater than the initial opening angle
$\theta_{\rm jet}^{\rm init}$ of each jet separately.  They find that if their angular separation
is less than this, the string endpoint and the kink in the gravitational theory
describe a single jet with substructure, not two jets.  In particular,
Casalderrey-Solana and Ficnar focus on identifying the smallest possible angular
separation between two resolved jets for a quark jet and a gluon jet with a particular summed energy.  
Since resolving the two jets requires them to be separated by more than their
individual initial opening angles, minimizing their separation means minimizing
their individual opening angles $\theta_{\rm jet}^{\rm init}$ in (\ref{eq:openingangle}).  
This means that the jets must be prepared with the minimum
possible value of $E_0$ or, equivalently from (\ref{CurlyC}), the maximum possible value of $\psi_1$.
Noting that the minimum possible value of $E_0$ is proportional to $T\sqrt{\lambda}$, 
the parametric dependence of $\theta_{\rm jet}^{\rm init}$ in (\ref{eq:openingangle}) reproduces
the parametric dependence of the smallest possible angular separation between two resolved jets identified in the
calculations reported in Ref.~\cite{Casalderrey-Solana:2015tas}.
}

\begin{acknowledgments}
We acknowledge helpful discussions with Jorge Casalderrey-Solana, Doga Gulhan, {Peter Jacobs}, Kristan Jensen, Yen-Jie Lee, Guilherme Milhano,
Daniel Pablos, Andrey Sadofyev, {Konrad Tywoniuk}, Wilke van der Schee and Korinna Zapp, and in particular thank
Guilherme Milhano and Korinna Zapp for a discussion of their work in progress. 
KR is grateful to the CERN Theory Division for hospitality as this research was being done.
The work of PC was  supported by the Fundamental
Laws Initiative at Harvard.
The work of KR was supported by the U.S.~Department of Energy
under Contract Number DE-SC0011090.
\end{acknowledgments}

\appendix

\section{The jet mass}
\label{app:JetMass}

We define the initial jet mass $M_{\rm init}$ via the initial  energy and momentum of the jet:
\begin{equation}
M^2_{\rm init} \equiv E^2_{\rm init} - \bm P_{\rm init}^2,
\end{equation}
where the jet energy and momentum are
\begin{align}
\label{eq:jetmomentumdef}
&E_{\rm init} \equiv \int d^3 x \, \langle T^{00}_{\rm jet} \rangle ,&
P^i_{\rm init} \equiv \int d^3 x \, \langle T^{0i}_{\rm jet} \rangle ,
\end{align}
where both integrals are to be evaluated at a time that is $\gg 1/T$ and $\ll x_{\rm therm}$.
Writing $| \bm P_{\rm init}| = E_{\rm init} - \frac{1}{2} \Delta$ and assuming $\Delta \ll E_{\rm init}$, the jet mass reads
\begin{equation}
\label{eq:jetmass2}
\frac{M_{\rm init}}{E_{\rm init}} = \sqrt{ \frac{\Delta}{E_{\rm init}}}.
\end{equation}

How does the initial jet mass scale with the opening angle $\theta_{\rm jet}^{\rm init}$ in the limit $\theta_{\rm jet}^{\rm init} \to 0$?
From (\ref{eq:thetajetinitresult}) we see that this question is equivalent to asking how the initial jet mass scales with the 
angle $\sigma_*$ in the $\sigma_* \to 0$ limit.  To this end, 
let us now express (\ref{eq:jetmass2}) in terms of the dual string variables.  
From the zero temperature results for the stress in Eqs.~(\ref{eq:particlestress2}) and (\ref{eq:DeltaStress}) and the definitions of the jet energy and momentum in 
(\ref{eq:jetmomentumdef}), it is easy to see that 
\begin{align}
&E_{\rm init} = - \int d\sigma \, \Pi^\tau_0(\sigma),&
|\bm P_{\rm init}| = -\int d\sigma \, \Pi^\tau_0(\sigma)/\xi(\sigma),
\end{align}
from which we can write
\begin{equation}
\label{eq:DeltaResult}
\Delta = - 2 \int d\sigma \, \Pi^\tau_0(\sigma)\left ( 1- 1/\xi(\sigma) \right ) = - 2 \int d\sigma \, \Pi^\tau_0(\sigma)\left ( 1- \cos \sigma \right ),
\end{equation}
where in the last line we have used (\ref{eq:xidef}).

From (\ref{eq:DeltaResult}) and the near-endpoint behavior of $\Pi^\tau_0$ in (\ref{eq:Pi00UV}), we see that $\Delta$ is finite in the $\sigma_* \to 0$ limit.
In contrast, the initial jet energy in (\ref{eq:initialenergy}) diverges like $1/\sigma_*^{3/2}$.  We therefore conclude $M_{\rm init}/E_{\rm init} \sim \sigma _*^{3/4}$,
or equivalently from (\ref{eq:thetajetinitresult}),
\begin{equation}
\frac{M_{\rm init}}{E_{\rm init}} = {\cal O}\left( \left(\theta^{\rm init}_{\rm jet}\right)^{3/4}\right)\ ,
\label{eq:jetmassresultApp}
\end{equation}
a result that we quoted in Eq.~(\ref{eq:jetmassresult0}).

The origin of the peculiar scaling (\ref{eq:jetmassresultApp}) is that there is energy flowing at large angles, or more precisely at angles
that are large compared to $\theta_{\rm jet}^{\rm init}$ in the  $\theta_{\rm jet}^{\rm init}\rightarrow 0$ limit,
and although this energy is not significant enough to have much of an affect on $\theta_{\rm jet}$ as we have 
defined it via $\bar x_\perp$, the half width at half maximum of $\Phi(x,x_\perp)$ illustrated in Figs.~\ref{fig:NormalizedPhi} and 
\ref{fig:TransverseEnvelope},
it does increase $E_{\rm init}$ sufficiently to yield the scaling (\ref{eq:jetmassresultApp}).
This is the reason why throughout this paper we have used the definition of the opening angle
that we have used, a definition that is analogous to defining it from the half width at half maximum of the
jet shape,
rather than attempting to define an opening angle via the jet mass.  Of course, using a definition based
upon, say, the half width at 10\% maximum of $\Phi(x,x_\perp)$ would be just as good.
But, as we noted in Section~\ref{sec:openingangle},
moments of $\Phi(x,x_\perp)$ like for example $\langle x_\perp \rangle$ are 
controlled by 
the large-$x_\perp$ tails of the distribution $\Phi(x,x_\perp)$, and so cannot be used.

\bibliographystyle{JHEP}
\bibliography{refs}

\end{document}